\documentclass[iop]{emulateapj}

\usepackage{amsmath}
\usepackage{graphicx,epsfig,epsf,graphics,epstopdf}
\usepackage{natbib,hyperref}
\def\apj{ApJ}
\def\nat{Nature}

\def\apjl{ApJL}
\def\apjs{ApJS}

\def\aap{A\&A}

\def\teff{T_\mathrm{eff}}

\shorttitle{Spectra of Earth-like Planets Through Geological Evolution Around FGKM Stars}
\shortauthors{Rugheimer et al.}

\begin{document}


\title{Spectra of Earth-like Planets Through Geological Evolution Around FGKM Stars }
\author{S. Rugheimer\altaffilmark{1,2,3} and L. Kaltenegger\altaffilmark{3}}

\submitted{Submitted February 25th, 2016. Accepted to ApJ December 27, 2017} 

\altaffiltext{1}{Centre for Exoplanet Science, University of St. Andrews, School of Earth and Environmental Sciences, Irvine Building, North Street, St. Andrews, KY16 9AL, UK}
\altaffiltext{2}{Harvard Smithsonian Center for Astrophysics, 60 Garden St., Cambridge, MA 02138 USA}
\altaffiltext{3}{Carl Sagan Institute, Department of Astronomy, Cornell University, Ithaca, NY 14853 USA}


\begin{abstract}

\setcounter{table}{0}
\renewcommand{\thetable}{5.\arabic{table}}

Future observations of terrestrial exoplanet atmospheres will occur for planets at different stages of geological evolution. We expect to observe a wide variety of atmospheres and planets with alternative evolutionary paths, with some planets resembling Earth at different epochs. For an Earth-like atmospheric time trajectory, we simulate planets from prebiotic to current atmosphere based on geological data. We use a stellar grid F0V to M8V ($\teff$ = 7000$\mskip3mu$K to 2400$\mskip3mu$K) to model four geological epochs of Earth's history corresponding to a prebiotic world (3.9$\mskip3mu$Ga), the rise of oxygen at 2.0$\mskip3mu$Ga and at 0.8$\mskip3mu$Ga, and the modern Earth. We show the VIS - IR spectral features, with a focus on biosignatures through geological time for this grid of Sun-like host stars and the effect of clouds on their spectra. 

We find that the observability of biosignature gases reduces with increasing cloud cover and increases with planetary age. The observability of the visible O$_2$ feature for lower concentrations will partly depend on clouds, which while slightly reducing the feature increase the overall reflectivity thus the detectable flux of a planet. The depth of the IR ozone feature contributes substantially to the opacity at lower oxygen concentrations especially for the high near-UV stellar environments around F stars. Our results are a grid of model spectra for atmospheres representative of Earth's geological history to inform future observations and instrument design and are publicly available online.

\end{abstract}

\section{INTRODUCTION}

\renewcommand{\arraystretch}{0.6}

Thousands of extrasolar planets have been found to date, some consistent with rocky planet models in the Habitable Zone \citep[see e.g.][]{torres2015, quintana2014, borucki2013, kaltenegger2013, batalha2013, borucki2011, kaltenegger2011, udry2007}. The occurrence rate for habitable Earth-like planets around M dwarfs is around 30\% \citep{dressing2015} and between 5-20\% around FGK stars \citep{burke2015, petigura2013, foreman2014, silburt2015}, indicating we will likely find a potentially habitable planet within 6$\mskip3mu$pc around a FGK star, and 2$\mskip3mu$pc around an M star\footnote{Note: the distances will be farther for transiting planets. These numbers are based on the RECONS list of nearby stars within 10$\mskip3mu$pc from www.recons.org}. TESS, NASA's explorer mission scheduled for launch in 2018, will search the whole sky for exoplanets around the closest and brightest stars, and will be sensitive to potentially habitable planets around M and K stars \citep{ricker2014}.

Current missions are unable to characterize terrestrial HZ planets, but future missions are in design to characterize the atmospheres of Earth-like planets \citep[e.g.][]{lawson2014, beichman2006, cash2006, traub2006}. For transiting terrestrial planets, JWST \citep[see][]{gardner2006,deming2009,kaltenegger2009} along with future ground- and space-based telescopes such as the E-ELT and the proposed High-Definition Space Telescope (HDST) or Large UV/Optical/Infrared Surveyor (LUVOIR) \citep{snellen2013, rodler2014, stark2014} will search for biosignatures in terrestrial exoplanet spectra. Since only a fraction of planets transit, direct imaging will maximize our chances of finding life with technology development currently underway for missions like the Starsade (Exo-S) and the Coronograph (Exo-C) mission concepts \citep{lawson2014, seager2014}, to characterize a planet through emergent visible and infrared spectra. 

Undoubtedly, we will encounter a diversity of terrestrial exoplanet atmospheres. Earth's own atmosphere has undergone significant evolution since formation \citep{walker1977}. Previous work from one of the authors examined Earth's spectral features through geological time \citep{kaltenegger2007}. Here we expand on this work by modeling these planets around an extensive grid of Sun-like stars (F0V - M8V) using a self-consistent climate and photochemistry model and a solar evolution model \citep{claire2012}. We focus on four geological epochs corresponding to Earth-like atmospheres at 3.9$\mskip3mu$Ga\footnote{Ga - billion years ago} (prebiotic), 2.0$\mskip3mu$Ga (right after the GOE\footnote{Great Oxygenation Event}), 0.8$\mskip3mu$Ga (after the NOE,\footnote{Neoproterozoic Oxygenation Event} a further increase in oxygen, and the start of multicellular life), and the modern Earth atmosphere. We model each epoch around stars with main sequence lifetimes longer than 2 billion years, used here as an estimate of the time needed for life to originate and evolve to a complexity that produces oxygen in sufficient quantities to accumulate in the atmosphere. Note this number is based on our own planet's history and the evolutionary timescales on other planets will likely vary. 

We create a grid of model spectra for atmospheres representative of Earth's geological history for a wide range of Sun-like host stars to interpret future observations and to inform instrument design. Exploring under which planetary conditions models that include biota show biosignatures that are clear versus ambiguous for different stellar types is a vital step.

In \S2, we describe our model and in \S3 we report the climate and photochemistry results of an Earth-like planet for 12 stellar types and the 4 different atmosphere models through geological time. In \S4, we examine the remote observability of spectral features in the VIS to IR, including biosignatures, and in \S5 and \S6, we conclude by summarizing the results and discussing their implications. 

\section{MODEL DESCRIPTION}

\subsection{Stellar and Planetary Models}

Our host star grid covers F0V to M8V ($\teff$ = 7000$\mskip3mu$K to 2400$\mskip3mu$K) \citep[see][for details on FKGM stellar models]{rugheimer2013, rugheimer2015b}. All stellar models use IUE\footnote{http://archive.stsci.edu/iue} or HST \citep{france2013} observations up to 3000$\mskip3mu$\AA\ combined with PHOENIX \citep{allard2014, allard2000} or ATLAS \citep{kurucz1979} stellar models for longer wavelengths.\footnote{The M1V, M3V and M8V are the active stellar models defined in \citet{rugheimer2015b}. Three additional M dwarfs with HST UV data and reconstructed Ly-$\alpha$ are GJ 581 (M3V, $\teff$ = 3498$\mskip3mu$K), GJ 832  (M1.5V, $\teff$ = 3620$\mskip3mu$K), and GJ 1214 (M4.5V, $\teff$ = 3250$\mskip3mu$K) \citep{france2013}. Though GJ 581, GJ 832 and GJ1214 planet models are not shown in figures for clarity, their spectra are available for download.}

Our model, EXO-Prime \citep{kaltenegger2010a}, is a coupled 1D radiative-convective atmosphere code. It has been developed for terrestrial exoplanets and comprises a 1D climate code \citep{kasting1986, pavlov2000, haqq2008}, a 1D photochemistry code \citep{pavlov2002, segura2005, segura2007}, and a 1D radiative transfer model \citep{traub1976, kaltenegger2009}. We use EXO-Prime to calculate the reflected light and thermal emission spectrum of a hypothetical Earth-like exoplanet through geological time orbiting stars spanning the FGKM stellar range.

For the oxygenated epochs at 2.0$\mskip3mu$Ga, 0.8$\mskip3mu$Ga and 0.0$\mskip3mu$Ga (modern Earth), we use a photochemical model including oxygen chemistry. The dominant contributions to emission spectra from Earth-like planet atmospheres come from the atmosphere below 60$\mskip3mu$km, and so we model our atmospheres to approximately 60$\mskip3mu$km ($10^{-4}$$\mskip3mu$bar) with 100 layers. The radiative forcing of clouds is included by adjusting the surface albedo of the modern Earth-Sun system \citep[following][]{kasting1984, segura2003}. The photochemistry code uses a reverse-Euler method to solve a chemical network of 55 chemical species linked by 220 reactions \cite[see][and references therein]{segura2010}. 

For the 3.9$\mskip3mu$Ga non-oxygenated epoch, we implement a 1D climate and photochemical model for high-CO$_2$/high-CH$_4$ terrestrial atmospheres \citep[see][and references therein]{pavlov2001, kharecha2005, segura2007}. We converge the climate model first and then input the temperature and pressure profiles into a photochemical model with 73 chemical species involved in 359 reactions.

Both the high and low O$_2$ photochemical models are stationary, and run to chemical equilibrium.

The remotely detectable spectra are calculated with a line-by-line radiative transfer model developed for exoplanet transmission and emergent spectra \citep{kaltenegger2007, kaltenegger2009, kaltenegger2010, kaltenegger2010a, kaltenegger2013}. The model computes absorption using the 2016 HITRAN database \citep{gordon2017} for the following species: H$_2$O, CO$_2$, N$_2$O, SO$_2$, H$_2$S, H$_2$O$_2$, ClO, HOCl, HO$_2$, H$_2$CO, HCl, CH$_3$Cl, NO$_2$, NO, OH, HNO$_3$, O$_3$, CH$_4$, O$_2$, and CO. We use cross-sections for N$_2$O$_5$ \citep{wagner2003} and CFCl$_3$ \citep{sharpe2004}. CO$_2$ line mixing becomes important especially for high CO$_2$ atmospheres and the method used is described in \citep{niro2005a, niro2005b}. For H$_2$O, CO$_2$, and N$_2$, we replace these far wings of the line-by-line calculation with measured continua data in these regions \citep{traub2002}. 

The most spectroscopically significant molecules in the emergent flux for Earth-like planets are: H$_2$O, CO$_2$, O$_2$, O$_3$, CH$_4$, CO, N$_2$O, CH$_3$Cl, H$_2$CO, and H$_2$O$_2$. In high resolution, very small features from HNO$_3$, NO$_2$, NO, and SO$_2$ start to contribute to the emergent spectrum but would be impossible to detect for any of the atmospheres modeled here. However, for planets with different atmospheric composition, which depend on several factors like outgassing rates and redox budget, these and other gases could become spectroscopically active \citep[see e.g.][for possible DMS, DMDS, SO$_2$, and H$_2$S spectral features]{domagal2011, kaltenegger2010a}. Note that many molecules not mentioned in the list above are critical to the photochemistry of these atmospheres and are included in the photochemical models described above to calculate the model atmospheres.

We calculate the spectrum at high spectral resolution, 0.01$\mskip3mu$cm$^{-1}$, and report the output on a grid of 0.1$\mskip3mu$cm$^{-1}$. The figures are smoothed further with a triangular kernel to a resolving power of 150 in the IR and 800 in the VIS for relevance to future direct detection spectra such as LUVOIR/HDST. We ran a test at a higher resolution, 0.001$\mskip3mu$cm$^{-1}$, and found an average of 1.5\% difference for the atmosphere temperatures and pressures we consider here in our reflectance and emission spectra. For a detailed analysis of the impact of resolution on radiative transfer calculations see \citet{hedges2016}. We previously validated EXO-Prime from the VIS to the IR using observations of Earth as an exoplanet from EPOXI, Mars Global Surveyor, Shuttle data, and multiple earthshine observations \citep{kaltenegger2007, kaltenegger2009, rugheimer2013}.

The geophysical community is divided on the rate and formation of continental crust \citep{arndt2013} though most records indicate an onset of plate tectonics by 3.0$\mskip3mu$Ga \citep{korenaga2013}. In absence of clear data, for all epochs we assume 70\% ocean, 2\% coast, and 28\% land, the present value for Earth. Before widespread vegetation (all epochs except modern Earth in our model), we adopt land surface compositions of 35\% basalt, 40\% granite, 15\% snow, and 10\% sand. For modern Earth we adopt 30\% grass, 30\% trees, 9\% granite, 9\% basalt, 15\% snow, and 7\% sand which includes the vegetation red edge (VRE) \citep[following][]{kaltenegger2007}. Surface reflectivities are from the USGS Digital Spectral Library\footnote{http://speclab.cr.usgs.gov/spectral-lib.html} and the ASTER Spectral Library\footnote{http://speclib.jpl.nasa.gov} \citep{baldridge2009}.

Clouds impact the detectability of atmospheric species. Water clouds increase the reflectivity in the VIS to NIR, though limit access to the lower atmosphere. In the IR, Earth-like clouds slightly decrease the overall emitted flux because they radiate at lower temperatures and can either decrease or increase the absorption features. The impact of clouds on the spectra is represented in our models by inserting continuum-absorbing/emitting layers at the altitudes of the clouds with the cloud fractions and heights based on modern Earth cloud data with the final spectrum represented by a weighted sum of the spectra for each cloud layer and the surface. We model here Earth-analogue clouds for all epochs comprising a 60\% global cloud cover divided between three layers: 40\% water clouds at 1$\mskip3mu$km, 40\% water clouds at 6$\mskip3mu$km, and 20\% ice clouds at 12$\mskip3mu$km \citep[following][]{kaltenegger2007} consistent with an averaged Earth cloud model. For a comparison of our Earth-analogue cloud implementation to remote sensing observations from EPOXI and to clear sky spectra see \citet{rugheimer2013}. We do not include aerosols in this model.

No noise has been included in the spectra to provide theoretical model input spectra for several instrument simulators. Instrument dependent noise needs to be added for specific instrument simulators to model realistic output. 

\subsection{Simulation Set-Up}

We focus on four geological epochs from Earth's history with atmosphere models defined in \citet{rugheimer2015a} \citep[following][]{kaltenegger2007}. We use a 1$\mskip3mu$bar atmosphere for all geological epochs modeled since geological evidence is consistent with paleo-pressures near modern values \citep{som2012, marty2013}. Some recent data has suggested atmospheric pressures lower than 0.5 bar in the late Archean \citep{som2016}, though this still remains an open question since the method of obtaining constraints from basalt vesicularity \citep{sahagian1994, sahagian2002a} is in practice confounded by vesicularity profiles that do not conform to the theory and by potential inflation of lava flows \citep{bondre2003, hon1994}.

For the Earth-Sun simulations, we use a solar evolution model for each epoch from \citet{claire2012}. In the absence of a self-consistent exobiology evolution model, we reduce the flux of all stellar types by the same decrease in luminosity Earth received at each geological time. This procedure is not meant to represent consistent stellar evolution across FGKM stars. Rather, we directly compare planets that receive the same bolometric flux across different stellar hosts with atmospheric compositions and biogenic fluxes modeled after Earth's evolution. M stars would not evolve meaningfully on the main sequence for these timescales and thus reducing the stellar flux by the same amount as in Earth's history is equivalent to observing a similar planet at an increased distance from its host star.

\begin{table}[h!]
\begin{center}
\caption{Surface Mixing Ratios over Geological Time for Earth-Sun \label{tableinitialmixingratios}}
\begin{tabular}{cccccccc}
\tableline\tableline
 & \multicolumn{5}{c}{Initial Mixing Ratios} \\
Age (Ga)& CO$_2$ & CH$_4$   & O$_2$ & O$_3$  & N$_2$O    \\
 \hline

3.9$\mskip3mu$Ga & 1.00E-01 & 1.65E-06 &1.00E-13 &  2.55E-19 & 0   \\
2.0$\mskip3mu$Ga & 1.00E-02 & 1.65E-03 & 2.10E-03 &  7.38E-09  & 8.37E-09\\
0.8$\mskip3mu$Ga & 1.00E-02 & 4.15E-04 & 2.10E-02 &  2.02E-08 & 9.15E-08\\
0.0$\mskip3mu$Ga & 3.55E-04 & 1.60E-06 & 2.10E-01 &  2.41E-08 & 3.00E-07\\
\hline   
\tableline
\tableline
\end{tabular}
\end{center}
\end{table}

The first epoch represents a prebiotic world, similar to early Earth (3.9$\mskip3mu$Ga) with a CO$_2$-rich atmosphere. This atmosphere model has a fixed mixing ratio of CO$_2$ = 0.1 and CH$_4$ =  $1.65 \times 10^{-6}$. We use an early Sun model at 3.9$\mskip3mu$Ga for the Earth-Sun case and reduce all other stars by a flux factor of 0.746, following the early Sun model. 

For the next three oxygenated epochs, we use the biological fluxes from \citet{rugheimer2015a, rugheimer2013} calculated from the Earth-Sun mixing ratios in Table \ref{tableinitialmixingratios}. Mixing ratios for CH$_4$ and N$_2$O are given in Table \ref{tablemixingratios} for each epoch for the grid of host stars from F0V to M8V. 

The second epoch corresponds to paleoproterozoic Earth (2.0$\mskip3mu$Ga), when oxygen started to rise in Earth's atmosphere. We use fixed surface mixing ratios of CO$_2$ = 0.01 and O$_2$ = $2.1 \times 10^{-3}$ (1\%$\mskip3mu$PAL, present atmospheric level). The biological fluxes for CH$_4$, N$_2$O and CH$_3$Cl are calculated for the Sun-Earth system at 2.0$\mskip3mu$Ga and then used as input for all other stellar types. We updated the biological fluxes for this epoch to CH$_4$ = $8.94 \times 10^{16}$$\mskip3mu$g$\mskip3mu$yr$^{-1}$, N$_2$O =  $3.45 \times 10^{13}$$\mskip3mu$g$\mskip3mu$yr$^{-1}$, and CH$_3$Cl = $8.18 \times 10^{11}$$\mskip3mu$g$\mskip3mu$yr$^{-1}$ due to an error in the calculated rates in \citet{rugheimer2015a} with the updated values being approximately 10\% of the old ones. With these new rates, we used a fixed mixing ratio for CH$_4$ for the planet orbiting the M8V and all MUSCLES host stars following \citet{rugheimer2015a} of $4.0 \times 10^{-3}$. For H$_2$ and CO, we used a fixed deposition velocity of 1.2 x 10$^{-4}$$\mskip3mu$cm$\mskip3mu$s$^{-1}$ and 2.4 x 10$^{-4}$$\mskip3mu$cm$\mskip3mu$s$^{-1}$ respectively \citep[following][]{domagal2014}. 

The third epoch corresponds to the proliferation of multicellular life on neoproterozoic Earth (0.8$\mskip3mu$Ga) when oxygen had risen to 10\%$\mskip3mu$PAL. We use fixed surface mixing ratios of CO$_2$ = 0.01 and O$_2$ = $2.1 \times 10^{-2}$. The biological fluxes for CH$_4$, N$_2$O and CH$_3$Cl are calculated for the Sun-Earth system at 0.8$\mskip3mu$Ga and then used as input for all other stellar types. H$_2$ and CO were set as in 2.0$\mskip3mu$Ga.

The fourth geological epoch corresponds to modern Earth \citep[simulations defined in][]{rugheimer2013}. This atmosphere has a fixed mixing ratio of CO$_2$ = 355$\mskip3mu$ppm and O$_2$ = 0.21 and biological fluxes corresponding to current Earth.

\begin{table*}[ht!]
\begin{center}
\caption{Surface Mixing Ratios for CH$_4$ and N$_2$O \label{tablemixingratios}}
\begin{tabular}{llcllllllllllllll}
\tableline\tableline
& \multicolumn{8}{c}{Mixing Ratios} \\
 \hline
 Star  & \multicolumn{2}{c}{3.9$\mskip3mu$Ga} & \multicolumn{2}{c}{2.0$\mskip3mu$Ga} &   \multicolumn{2}{c}{0.8$\mskip3mu$Ga} & \multicolumn{2}{c}{Modern Earth}   \\
Type    & CH$_4$ & N$_2$O   & CH$_4$ & N$_2$O  & CH$_4$ & N$_2$O   & CH$_4$ & N$_2$O  \\

  \hline   
\tableline
F0V        &  1.65E-06  &   -  & 1.24E-03  &   8.26E-09  &   5.65E-04   &    1.10E-07  &    3.58E-06  &   1.63E-07   \\
F7V        &  1.65E-06  &   -  &  1.29E-03  &  5.75E-09  &   5.68E-04   &   1.01E-07    &   2.46E-06  &    2.98E-07   \\
Sun        &  1.65E-06  &  -  &  1.65E-03   &  8.37E-09  &   4.15E-04   &   9.15E-08    &   1.73E-06  &    3.03E-07   \\
G8V       &  1.65E-06  &  -  &  1.72E-03   &  2.47E-08  &   4.60E-04   &   6.64E-08    &   4.20E-04  &    3.08E-07   \\
K2V        &  1.65E-06 &  -  &  4.12E-03   &  3.47E-08   &   1.14E-03   &   8.58E-08   &   4.53E-04  &    3.10E-07   \\
K7V        &  1.65E-06  & -   & 1.99E-02  &  4.93E-08  &  1.16E-02      &  1.60E-07    &   1.25E-04  &   8.40E-07   \\ 
M1V       &  1.65E-06 &  -  &  1.41E-02   &  3.90E-08  &   1.13E-02    &   1.31E-07    &   3.71E-04  &   7.04E-07   \\
M3V       &  1.65E-06  &  -  &  1.77E-02  &  4.42E-08  &   1.10E-02    &   1.14E-07    &    1.45E-04  &  8.18E-07   \\
M8V       &  1.65E-06 &  -  & 4.00E-03   &   3.37E-07  &   1.10E-02    &   8.46E-07  &   1.00E-03  &   3.09E-06   \\  
GJ 581   & 1.65E-06  & -   &  4.00E-03  &   7.65E-07  &    1.10E-02   &   1.81E-06  &    1.35E-03  &   4.92E-06   \\
GJ 832   & 1.65E-06  & -  & 4.00E-03  &   2.86E-06   &   1.10E-02     &   7.18E-06  &     5.46E-04  &   1.52E-05  \\
GJ 1214 & 1.65E-06  & -  & 4.00E-03   &  1.15E-07    &  1.10E-02    &   2.57E-07   &    1.63E-03  &    1.10E-06   \\
\tableline
\tableline
\end{tabular}
\end{center}
\end{table*}

\section{ATMOSPHERIC CLIMATE AND PHOTOCHEMISTRY RESULTS}

\begin{figure*}[ht!]
\centering
\includegraphics[scale=0.6,angle=0]{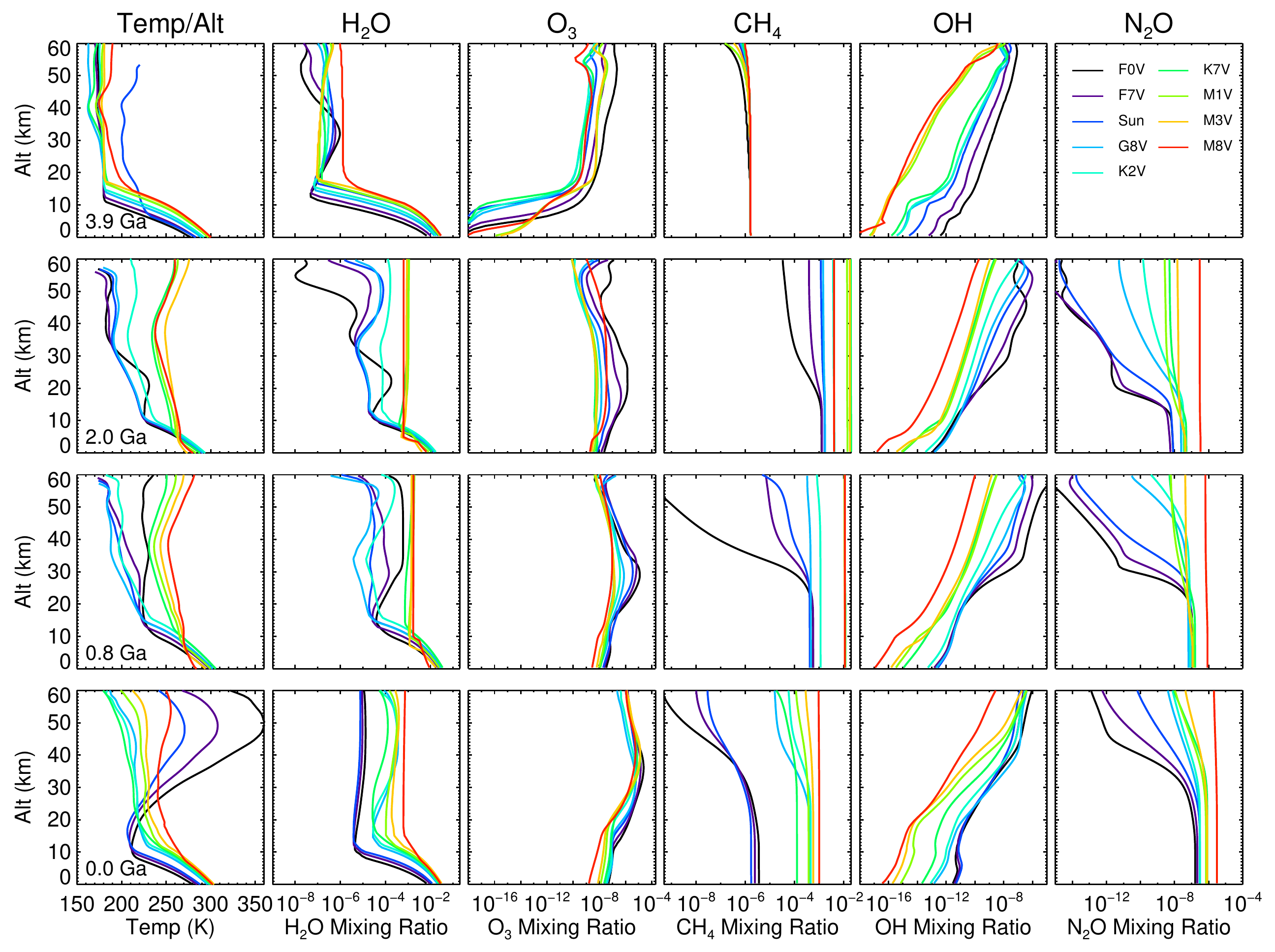}
\caption{Planetary temperature vs altitude profiles and mixing ratio profiles profiles for H$_2$O, O$_3$, CH$_4$, OH, and N$_2$O (left to right) for a planet orbiting the grid of FGKM stellar models with a prebiotic atmosphere corresponding to 3.9$\mskip3mu$Ga (1st row), the early rise of oxygen at 2.0$\mskip3mu$Ga (2nd row), the start of multicellular life on Earth at 0.8$\mskip3mu$Ga (3rd row), and the modern atmosphere (4th row).\label{tpchemsepochs}}
\end{figure*}

The temperature vs altitude atmospheric profile and the H$_2$O, O$_3$, CH$_4$, OH, and N$_2$O mixing ratio profiles for Earth-like atmosphere models around FGKM dwarfs the four epochs are shown in the columns of Fig. \ref{tpchemsepochs}. Each row corresponds to one of the four atmospheres modeled: 3.9$\mskip3mu$Ga, 2.0$\mskip3mu$Ga, 0.8$\mskip3mu$Ga and the modern atmosphere (0.0$\mskip3mu$Ga), respectively, and each color corresponds to one grid star. CH$_3$Cl profiles are not shown, but follow the same trends as CH$_4$. Since both O$_2$ and CO$_2$ are well mixed in the atmosphere, their vertical mixing ratio profiles are not shown, but are given in Table \ref{tableinitialmixingratios}.

In the first column of Fig. \ref{tpchemsepochs}, we present the temperature vs altitude profiles for the simulated planets and epochs. The temperature inversions for the earlier atmospheres are weaker than for modern Earth due to the lower ozone concentrations and thus less heating in the stratosphere. Within each atmosphere, the increased UV flux for hotter stars photolyzes more H$_2$O and CH$_4$, causing those stratospheres to be cooler due to less heating from those gases. This trend is reversed in the modern atmosphere where stratospheric heating by O$_3$ dominates.

\begin{figure}[h!]
\centering
\includegraphics[scale=0.37,angle=0]{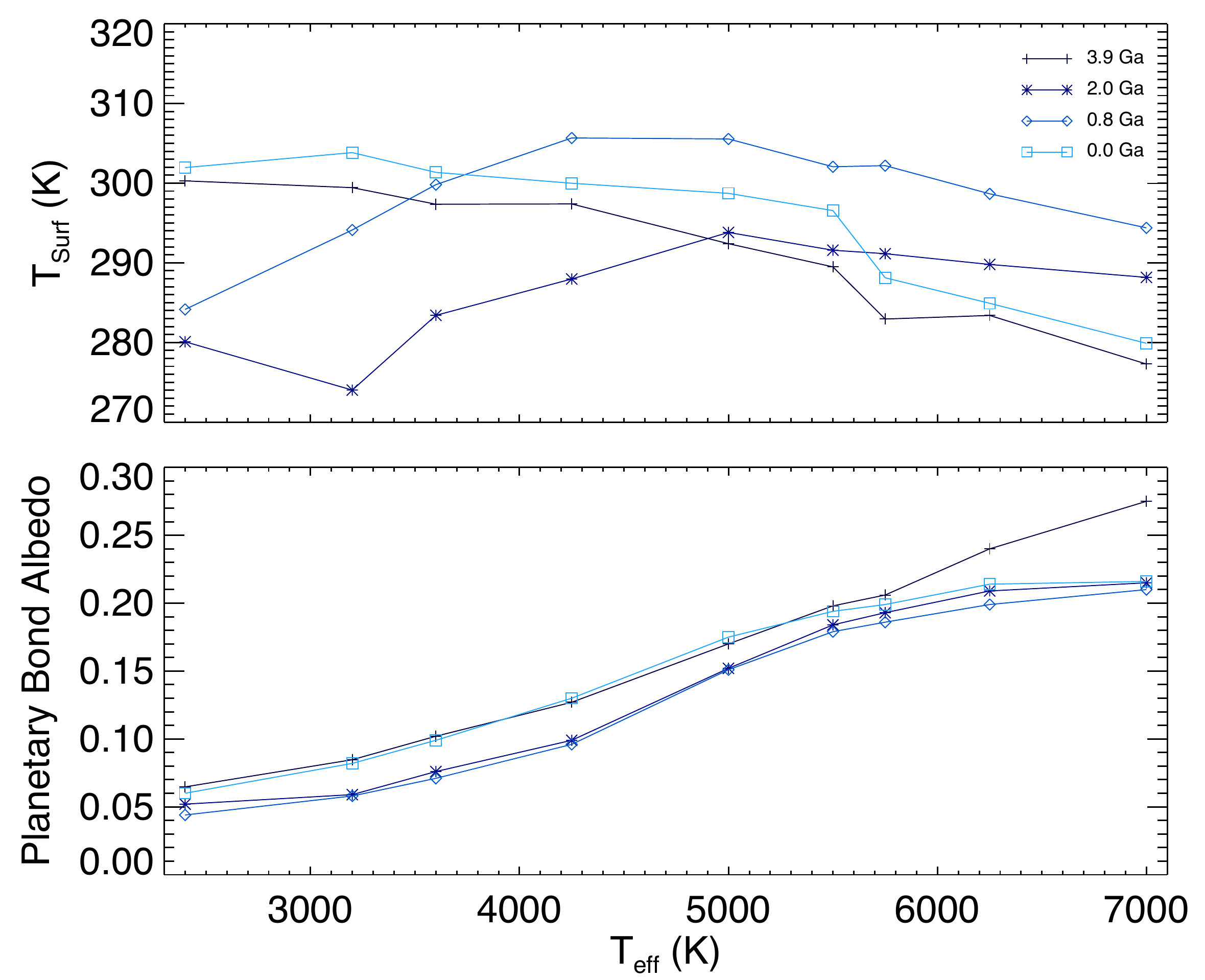}
\caption{Calculated surface temperatures (top) and planetary bond albedos (bottom) for each geological epoch and stellar type.\label{tempalbedo}}
\end{figure}

H$_2$O concentrations (Fig. \ref{tpchemsepochs}, 2nd column) in the troposphere are determined by the surface temperature (see Fig. \ref{tempalbedo}). H$_2$O is predominantly formed in the stratosphere by CH$_4$ via reaction R1.
\begin{align}\label{stratH2O}
\textrm{CH}_4\ + \textrm{OH} &\rightarrow \textrm{CH}_3 + \textrm{H}_2\textrm{O}  \tag{R1} 
\end{align}

H$_2$O is present in higher concentrations around cooler stars and by vertical transport from the troposphere in planets where there is little or no temperature inversion. H$_2$O increases for the middle epochs which have a higher greenhouse gas concentration and a higher stellar luminosity than for the prebiotic model. H$_2$O in the stratosphere is removed by photolysis or reactions with O($^1$D) (reactions R2-R3). 
\begin{align} \label{OH}
\textrm{H}_2\textrm{O} + h\nu &\rightarrow \textrm{H + OH}  \tag{R2} \\ 
\textrm{H}_2\textrm{O} + \textrm{O(}^1\textrm{D)} &\rightarrow \textrm{OH + OH} \tag{R3}
\end{align}

Both reactions produce OH and therefore lead to a higher OH concentration in all epochs for hotter stars and an increase with each epoch as the atmospheres become more oxygenated (Fig. \ref{tpchemsepochs}, 5th column). Photolysis of H$_2$O occurs for $\lambda < 2000$$\mskip3mu$\AA. For the prebiotic atmosphere there is much more flux at those wavelengths penetrating the atmosphere \citep[see][]{rugheimer2015a} and a corresponding decrease in stratospheric H$_2$O for that epoch. 

Ozone is produced primarily by photolysis of O$_2$ for $\lambda < 2400 $\AA\ via the Chapman reactions (reactions R4-R7) \citep{chapman1930}. 

\begin{align} \label{O3}
\textrm{O}_2 + h\nu &\rightarrow \textrm {O + O ($\lambda < $ 240 nm)} \tag{R4}  \\
\textrm{O + O$_2$ + M} &\rightarrow \textrm{O$_3$ + M} \tag{R5} \\
\textrm{O}_3 + h\nu &\rightarrow \textrm {O$_2$ + O ($\lambda < $ 320 nm)} \tag{R6}  \\
\textrm{O$_3$ + O} &\rightarrow \textrm{2 O$_2$} \tag{R7}
\end{align} 

Due to photolysis of other species such as CO$_2$ and H$_2$O in the prebiotic atmosphere at 3.9$\mskip3mu$Ga, there is enough free oxygen to form substantial O$_3$ in the stratosphere, although surface concentrations are up to 10 orders of magnitude lower. Already by 2.0$\mskip3mu$Ga there is sufficient O$_3$ for the F stars to create the characteristic ozone bulge in the mixing ratio profile which then increases further for the atmospheres at 0.8$\mskip3mu$Ga and modern Earth (Fig. \ref{tpchemsepochs}, 3rd column). 

Because O$_3$ is primarily produced by $\lambda < 2400\mskip3mu$\AA\ and destroyed by $\lambda < 3200\mskip3mu$\AA, the balance of UV incident flux shortwards and longwards of 2400$\mskip3mu$\AA\ sets the amount of O$_3$ in the atmosphere \citep[see also][]{segura2005,domagal2011}. Hotter stars tend to have more UV continuum flux, which for F, G, and early K stars dominates the shorter wavelengths, $\lambda < 2400\mskip3mu$\AA. But for the later K and all M spectral types, relatively little of the overall UV flux is contributed by the continuum (with most in the Ly-$\alpha$ line at 1216$\mskip3mu$\AA). Therefore, the higher UV flux at shorter versus longer wavelengths for K and M stars creates ozone, as seen in the increase in ozone column densities for the cooler stars.

CH$_4$ mixing ratio profiles are shown in column 4 of Fig. \ref{tpchemsepochs}. The largest sink for CH$_4$ in the troposphere and stratosphere is OH. OH concentrations increase as the atmosphere is oxidized. The biotic surface CH$_4$ flux in our model is greatest at 2.0$\mskip3mu$Ga and correspondingly shows the highest CH$_4$ atmospheric concentrations corresponding to increased CH$_4$ production from methanogens on Earth at that time and a longer atmospheric lifetime due to lower OH concentrations. For the F and G stars there is increased destruction of CH$_4$ in the upper atmosphere due to larger OH concentrations and a larger UV flux ($\lambda < 1500\mskip3mu$\AA). For the K and M stars, however, the highest CH$_4$ concentrations are found for the epoch with the second largest surface CH$_4$ flux at 0.8$\mskip3mu$Ga (see also Table \ref{tablemixingratios}) due to higher stratospheric H$_2$O abundances from increased stellar luminosity and therefore slightly higher CH$_4$ concentrations (see R1).

N$_2$O concentrations increase in our models for each oxygenated epoch due to increasing biological flux like on Earth. The destruction of N$_2$O is driven by photolysis by $\lambda <$ 2200$\mskip3mu$\AA\ and therefore tracks the UV environment of the host star with lower N$_2$O concentrations around hotter stars. Note that we model no N$_2$O flux for our prebiotic atmosphere at 3.9$\mskip3mu$Ga. 

The calculated planetary Bond albedo (surface + atmosphere) and surface temperature for each stellar type and geological epoch are shown in Fig. \ref{tempalbedo}. The Bond albedo increases for hotter stars due to increased Rayleigh scattering at shorter wavelengths. The surface temperature is affected by a combination of factors. The higher albedos of planets orbiting hotter stars decrease the surface temperature. But also more efficient cooling in the stratosphere lowers the surface temperature. 

\section{RESULTS: SPECTRA OF EARTH-SIZED PLANETS THROUGH GEOLOGICAL TIME ORBITING FGKM STARS}

Spectra of Earth-sized planets orbiting FGKM stars through geological time show measurable differences in spectral feature depths \citep[see also][]{kaltenegger2007}. We assume full phase for all spectra presented to show the maximum observable flux. No noise has been added to these spectra to provide inputs for a wide variety of instrument simulators for both secondary eclipse and direct detection simulations.

\subsection{Earth-like Visible/Near-infrared Spectra (0.4$\mskip3mu\mu$m - 4$\mskip3mu\mu$m)}

Fig. \ref{VISspectra} shows the reflected visible/near-infrared spectra from 0.4 to 2$\mskip3mu\mu$m of Earth-like planets orbiting the FGKM grid stars through geological time modeled after Earth's evolution for four epochs. {Figs. \ref{oxygen} - \ref{visH2O} highlight the individual bands for O$_2$, O$_3$, H$_2$O and CH$_4$. In the VIS, the abundance of the species determines the depth of the absorption features. We assume Earth-analogue 60\% cloud cover and refer the reader to \citet{rugheimer2013} for comparison to clear sky. The high-resolution spectra have been calculated at 0.01$\mskip3mu$cm$^{-1}$ steps and then smoothed to a resolving power of 800 for clarity. The region 2-4$\mskip3mu\mu$m has low integrated flux levels and therefore is not shown in Fig. \ref{VISspectra} though the NIR is shown in Figs. \ref{IRfeaturesSun} to \ref{IRfeaturesM8AHiRes} in the appendix for the individual chemical features.

\begin{figure}[h!]
\centering
\includegraphics[scale=0.45,angle=0]{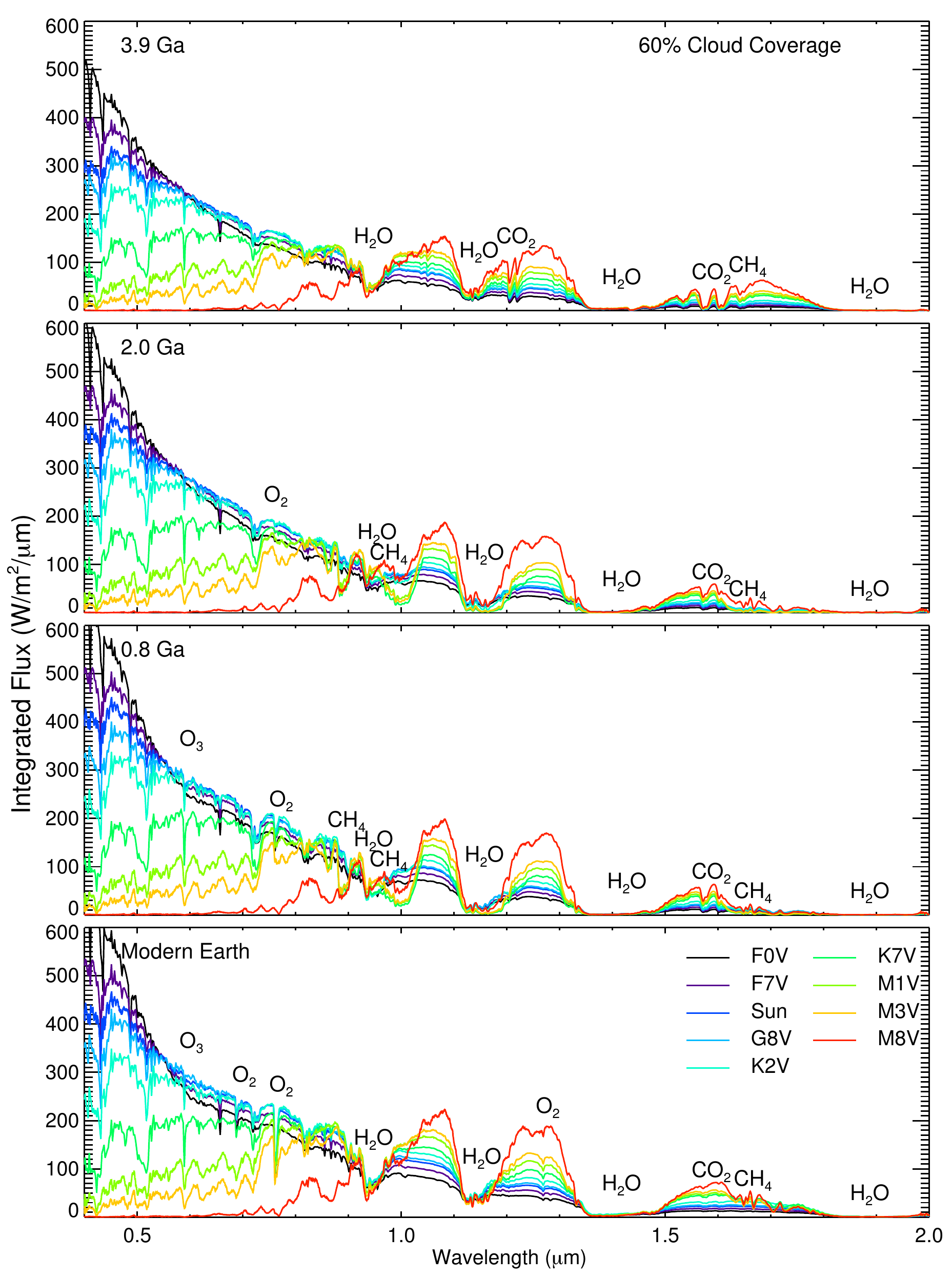}
\caption{Disk-integrated VIS/NIR spectra at a resolution of 800 at the TOA for an Earth-like planet for the grid of stellar and geological epoch models assuming 60\% Earth-analogue cloud coverage. For individual features highlighting the O$_2$, O$_3$, and H$_2$O/CH$_4$ bands in the VIS spectrum see Figs. \ref{oxygen}, \ref{visO3}, and \ref{visH2O}, respectively.} \label{VISspectra}

\end{figure}

Fig. \ref{VISspectra} shows that the host star has a strong influence on the exoplanet's spectrum and detectable features for several reasons. First, due to the increased stellar flux at shorter wavelengths for an F or G type star, Rayleigh scattering is much more pronounced for F and G stars than for K and M stars and therefore greatly increases the flux from 0.4 to 0.8$\mskip3mu\mu$m for an Earth-like planet around hotter stars. Second, late K and M dwarfs emit stronger in the NIR and thus the flux from 1-2$\mskip3mu\mu$m is higher around Earths orbiting M dwarfs than the Earth-Sun equivalent. Note that we scale the stellar luminosity to be the luminosity equivalent to the Sun at that geological epoch (as explained in \S2.2). Thus the earlier epochs have less absolute reflected light flux than the later epochs.

The most notable features in the VIS/NIR spectra (shown in Fig. \ref{VISspectra}) are the broad triangular O$_3$ feature from 0.45-0.74$\mskip3mu\mu$m (the Chappuis band), O$_2$ at 0.76$\mskip3mu\mu$m, H$_2$O at 0.95, 1.1, 1.4, and 1.9$\mskip3mu\mu$m, CO$_2$ at 1.2, 1.62$\mskip3mu\mu$m, and CH$_4$ at 0.6, 0.7, 0.8, 0.9, 1.0 and 1.7$\mskip3mu\mu$m that together can be used as a biosignature \citep{lederberg1965, lovelock1975, sagan1993}. Note that any shallow spectral features like the O$_3$ feature from 0.45-0.74$\mskip3mu\mu$m would require very high signal-to-noise to be detectable in low resolution. For FGK stars, features further into the NIR will be more difficult to detect in reflected light due to the low flux from the star at those wavelengths, however these features are shown for completeness in Figs. \ref{IRfeaturesSun} to \ref{IRfeaturesM8AHiRes} in the appendix. In particular, CO$_2$ has a strong feature at 2.0$\mskip3mu\mu$m and H$_2$O dominates much of the rest of the NIR spectrum.

Fig. \ref{oxygen} shows the details for one of the most notable feature in the VIS, the O$_2$ band at 0.76$\mskip3mu\mu$m in relative reflectivity and in reflected emergent flux for a clear sky and 60\% cloud cover model stars. In the prebiotic atmosphere, no oxygen feature is present in either the emergent flux or in relative reflectivity due to only minimal concentrations of photochemically produced oxygen at that time and assuming a similar redox state as Earth. As oxygen increases, the relative absorption depth similarly increases from 1\% (row 2), to 10\% (row 3), and finally to 100\% (row 4) PAL (Present Atmospheric Level) O$_2$. We also note that the O$_2$ feature becomes difficult to detect for the coolest M dwarf in our sample, the M8V, even in the modern atmosphere due to low intrinsic stellar luminosity at those wavelengths \citep[see also][]{rugheimer2015b}.

\begin{figure}[h!]
\centering
\includegraphics[scale=0.47,angle=0]{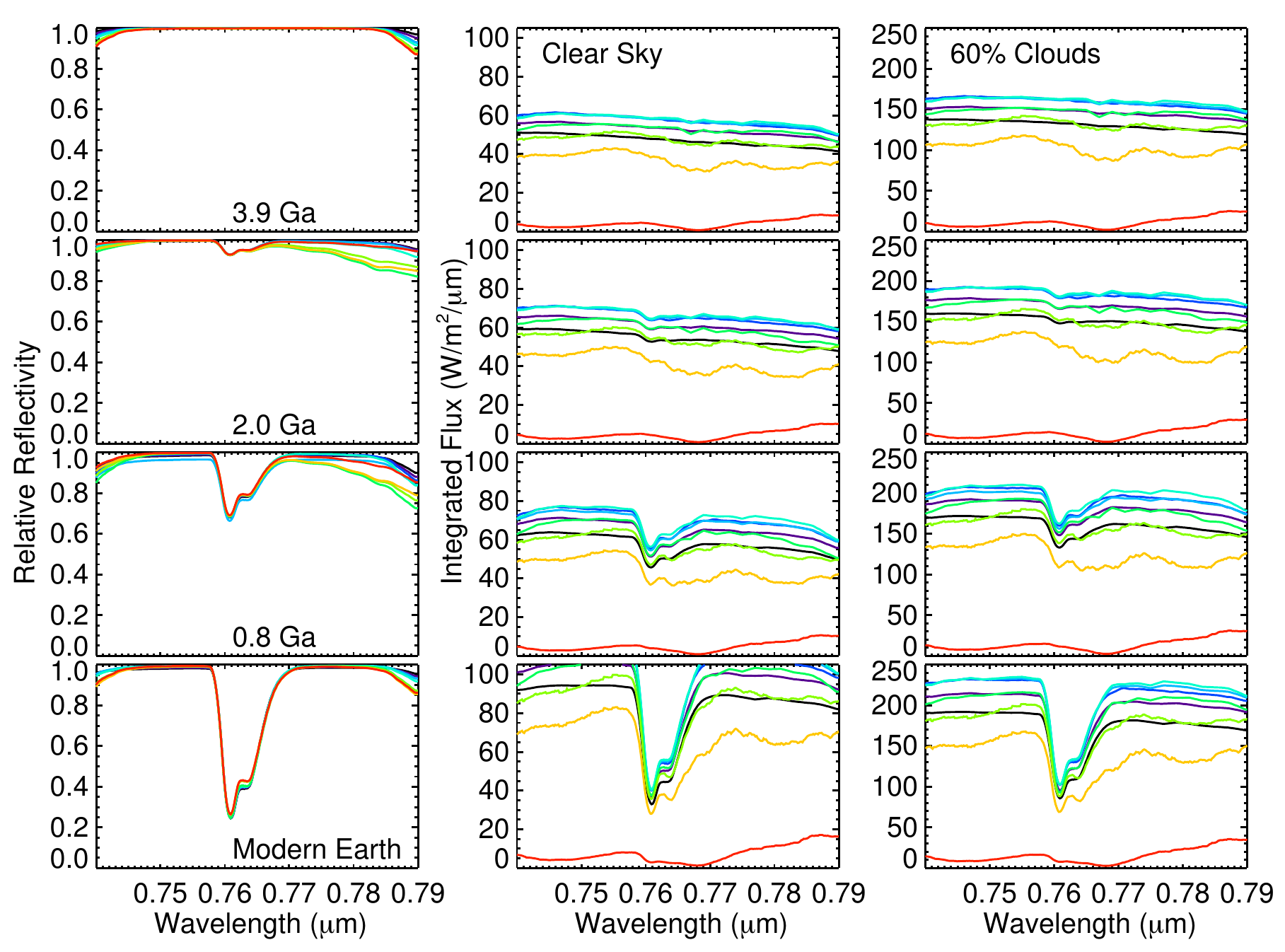}
\caption{Disk-integrated spectra (R = 800) of the O$_2$ feature at 0.76$\mskip3mu\mu$m for clear sky in relative reflectivity (left) and the detectable reflected emergent flux for clear sky (middle) and 60\% cloud cover (right). Coloring is the same as in Fig. \ref{VISspectra}. \label{oxygen}}
\end{figure}

\begin{figure*}[ht!]
\centering
\includegraphics[scale=0.5,angle=0]{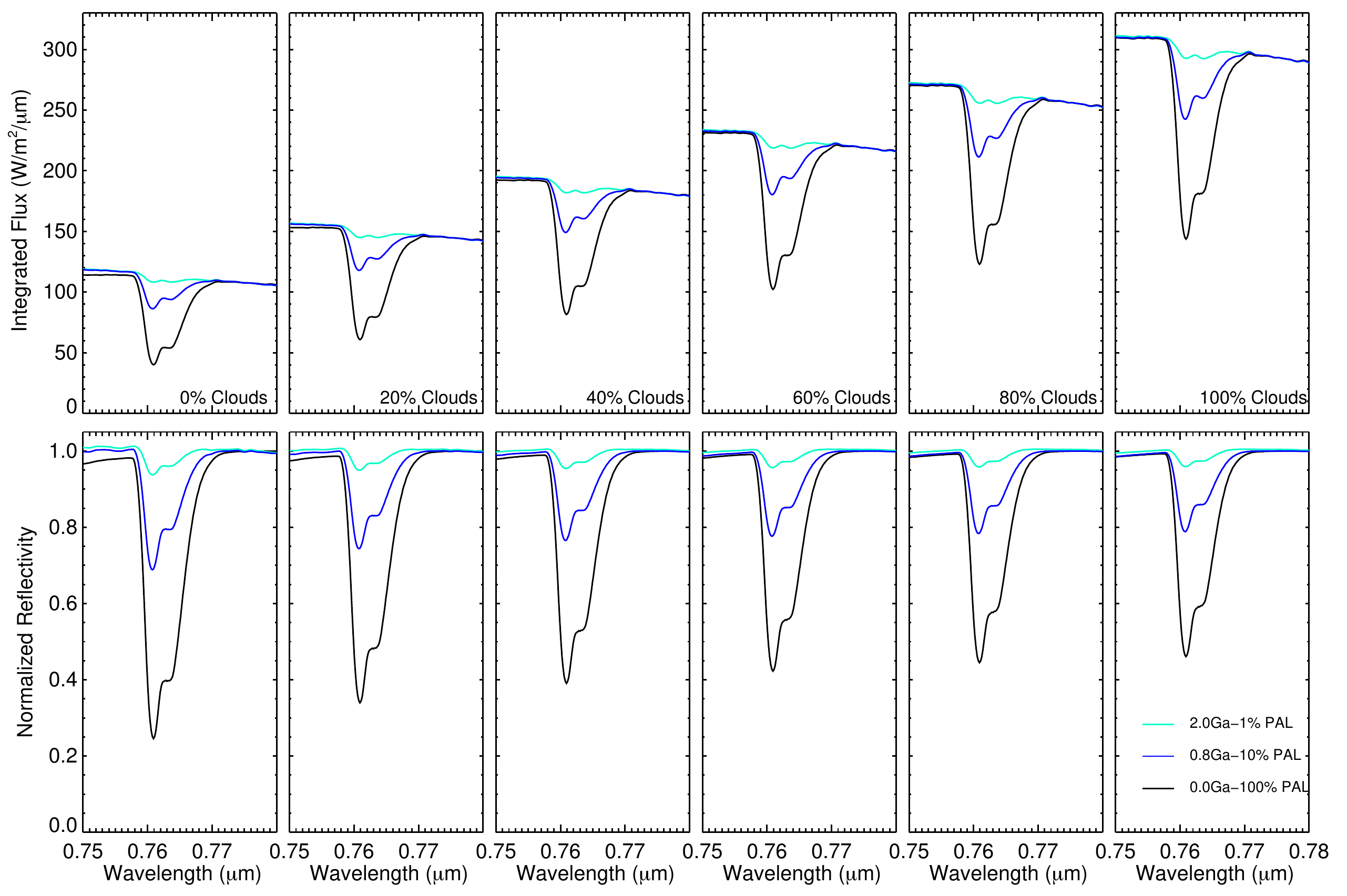}
\caption{O$_2$ feature at 0.76$\mskip3mu\mu$m in integrated flux (top) and normalized relative reflectivity (bottom) for 0\%, 20\%, 40\%, 60\%, 80\%, and 100\% cloud coverage (left to right) for the Earth-Sun case through geological time. The black lines represent modern PAL O$_2$ concentrations, the dark blue lines 10\%$\mskip3mu$PAL, and the light blue lines 1\%$\mskip3mu$PAL. Cloud coverage and surface reflectivity are included. The reflected stellar light and Rayleigh scattering effects have been removed. \label{oxygenclouds}}
\end{figure*}

To further clarify the dependence of the detectability O$_2$ A-band at 0.76$\mskip3mu\mu$m on cloud coverage, Fig. \ref{oxygenclouds} shows this feature for the three biotic epochs for 0\%, 20\%, 40\%, 60\%, 80\%, and 100\% cloud coverage for the Earth-Sun case with the relative fractions of 1$\mskip3mu$km, 6$\mskip3mu$km, and 12$\mskip3mu$km clouds kept the same as in the Earth-analogue model. In this figure, we show the integrated flux (top panel) and normalized reflectivity (bottom panel) including the effect of global cloud coverage and the surface reflectance, removing the impact of the reflected stellar light and Rayleigh scattering since in real observations of exoplanets it will be possible to remove the stellar spectrum from the integrated flux with contemporaneous observations of the host star and Rayleigh scattering may be approximated. No planets with a substantial atmosphere in our solar system have a cloud free atmosphere nor will we have detailed surface or cloud information in the first generation of terrestrial exoplanet characterization missions. Higher altitude clouds block more of the feature below them than lower altitude clouds in the reflected light spectrum, but as seen in Fig. \ref{oxygenclouds} this is in part balanced by the increased signal gained by a higher albedo from either low or high altitude clouds \citep[see also][]{tinetti2006b,kitzmann2011a}.

Fig. \ref{visO3} displays the individual feature for the Chappuis O$_3$ feature at 0.6$\mskip3mu\mu$m. The left panel shows the relative reflectivity. The O$_3$ visible feature depth is most pronounced for the hotter stellar types which have higher UV fluxes and therefore more O$_3$ production. The feature has a similar absorption depth at 0.8$\mskip3mu$Ga as in the modern atmosphere for the F and G stars. For the cooler K and M stars, the O$_3$ feature depth in the visible remains close to the absorption depth of the abiotic levels until the modern Earth atmosphere. In low resolution, the detection of the broad 0.6$\mskip3mu\mu$m O$_3$ feature in reflected light would require very high signal to noise to detect even for the highest ozone abundances for all stellar types.

Fig. \ref{visH2O} displays the H$_2$O feature at 0.95$\mskip3mu\mu$m and CH$_4$ features at 0.9 and 1.0$\mskip3mu\mu$m. In relative reflectivity, the prebiotic atmosphere at 3.9$\mskip3mu$Ga is most similar to the modern atmosphere in terms of absorption feature depth for H$_2$O and CH$_4$. In the reflected emergent spectra, the CH$_4$ will require high signal to noise to detect in the VIS/NIR. The surface reflectivity in the pre-vegetation atmospheres (3.9 - 0.8$\mskip3mu$Ga) also influences the detectability of features. Because of that, the H$_2$O feature shows the deepest absorption in the modern atmosphere despite having higher abundances at 2.0$\mskip3mu$Ga and 0.8$\mskip3mu$Ga. The relative absorption is still strongest in those middle epochs, while the seemingly increased depth in the modern epoch is due to the increased reflectance due to the vegetation red edge (VRE) starting at 0.7$\mskip3mu\mu$m, increasing the absolute flux level for the modern epoch compared to earlier epochs.
\begin{figure}[h!]
\centering
\includegraphics[scale=0.55,angle=0]{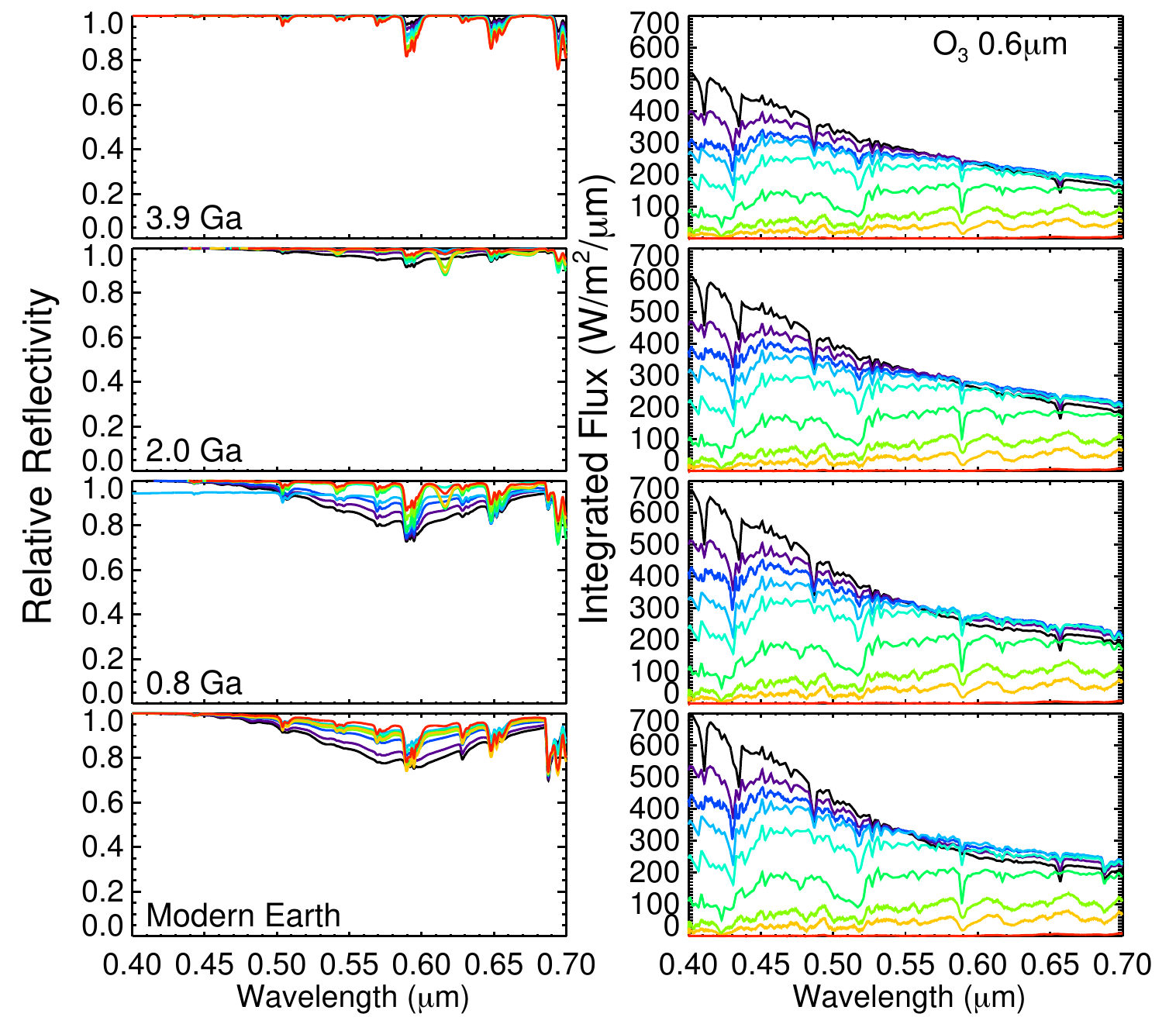}
\caption{Disk-integrated spectra (R = 800) of the broad O$_3$ Chappuis feature at 0.6$\mskip3mu\mu$m for clear sky in relative reflectivity (left) and in reflected emergent flux for a 60\% cloud cover (right). Coloring is same as in Fig. \ref{VISspectra}. \label{visO3}}
\end{figure}
\begin{figure}[h!]
\centering
\includegraphics[scale=0.55,angle=0]{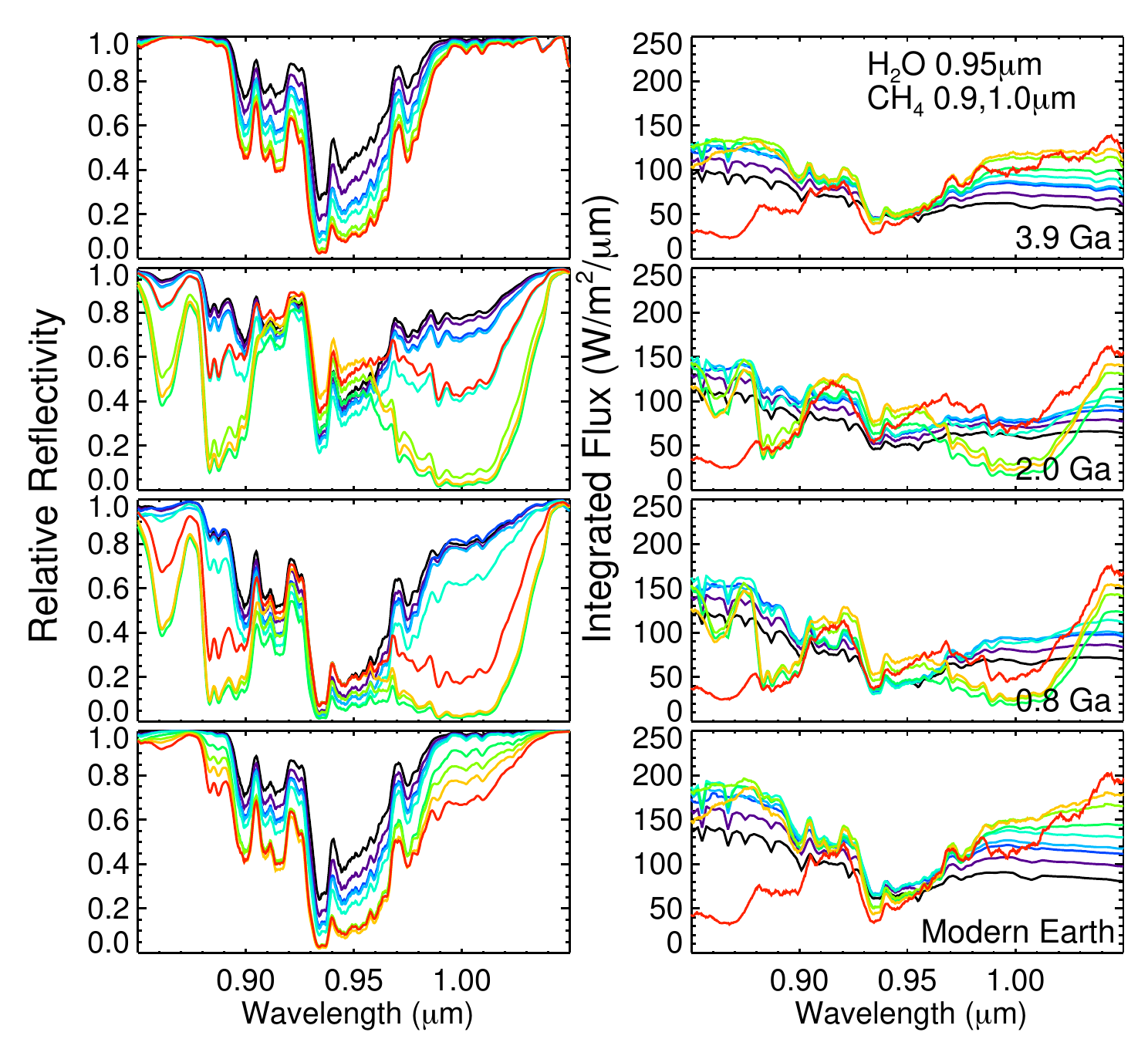}
\caption{Disk-integrated spectra (R = 800) of the H$_2$O feature at 0.95$\mskip3mu\mu$m and CH$_4$ features at 0.9 and 1.0$\mskip3mu\mu$m for clear sky in relative reflectivity (left) and in reflected emergent flux for a 60\% cloud cover (right). Coloring is same as in Fig. \ref{VISspectra}. \label{visH2O}}
\end{figure}

For all epochs and all host stellar types, H$_2$O, CO$_2$ and CH$_4$ are the dominant features in the visible and near-IR spectrum. As oxygen rises, O$_2$ and O$_3$ features increasingly contribute to the visible spectrum for all stellar types. For CO$_2$ atmospheres, CO has features in the VIS/NIR at 1.57$\mskip3mu\mu$m and 2.35$\mskip3mu\mu$m with the features reducing in depth for lower CO$_2$ concentrations. N$_2$O has many features in the NIR at 2.1-2.3$\mskip3mu\mu$m, 2.5-3$\mskip3mu\mu$m, and 3.5-4$\mskip3mu\mu$m and CH$_3$Cl has a feature at 3.35$\mskip3mu\mu$m if there is a biological source for these gases, which we model from 2.0 Ga onward. In addition, H$_2$CO has a strong feature at 3.56$\mskip3mu\mu$m for the middle two epochs where there is both ample CH$_4$ and enough O$_2$ to create species needed for its formation, but not too much OH for its destruction. CH$_3$Cl, H$_2$CO, CO, CH$_4$, and N$_2$O have stronger features in the spectra for planets around less UV active stars. 

However, even though H$_2$CO, N$_2$O, and CH$_3$Cl have strong features, they are all undetectable in the final NIR spectrum which is dominated by the much stronger and overlapping H$_2$O, CO$_2$ and CH$_4$ absorption features for any epoch or host star type. The 1.57$\mskip3mu\mu$m CO feature is likewise obscured by a larger a CO$_2$ feature at that wavelength. The only feature without significant overlap is the 2.35$\mskip3mu\mu$m CO feature for the abiotic high CO$_2$ atmosphere. See Figs. \ref{IRfeaturesSun} to \ref{IRfeaturesM8AHiRes} in the appendix for a plot of the individual features in the VIS/NIR for H$_2$O, CO$_2$, CH$_4$, O$_2$, O$_3$, CO, H$_2$CO, N$_2$O, and CH$_3$Cl for all epochs for the Sun and M8A host star case.

\subsection{Earth-like Infrared Spectra (4$\mskip3mu\mu$m - 20$\mskip3mu\mu$m)}

Fig. \ref{IRspectra} shows the thermal emission IR spectra from 4 to 20$\mskip3mu\mu$m of Earth-like planets orbiting the FGKM stars for four geological epochs modeled after Earth's evolution and assuming a 60\% cloud cover. Figs. \ref{ozone} - \ref{irCO2} highlight the individual bands for O$_3$, H$_2$O and CO$_2$. We refer the reader to \citet{rugheimer2013} for comparison to clear sky. In the IR, both the concentration and the temperature difference between the continuum and the emitting/absorbing layer influence the depth of absorption features. The IR spectra are displayed with a resolving power of 150.

\begin{figure}[h!]
\centering
\includegraphics[scale=0.45,angle=0]{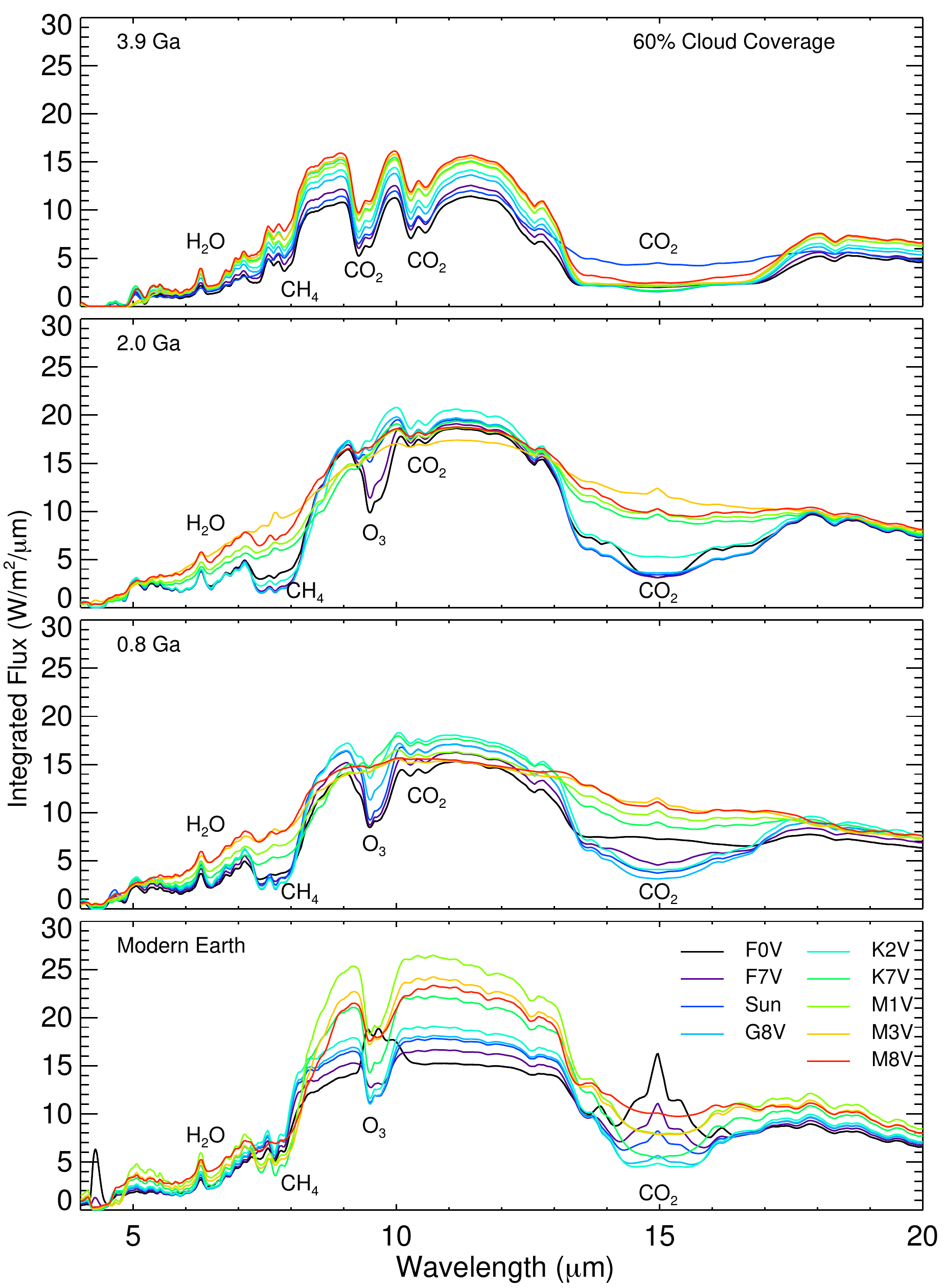}
\caption{Smoothed, disk-integrated IR spectra at the TOA for an Earth-like planet for the grid of stellar and geological epoch models assuming 60\% Earth-analogue cloud coverage. For individual features highlighting the O$_3$, H$_2$O/CH$_4$, and CO$_2$ bands in the IR spectrum see Figs. 9, 10, and 11, respectively.} \label{IRspectra}
\end{figure}

The carbon dioxide features at 9.3 and 10.4$\mskip3mu\mu$m and the ozone feature at 9.6$\mskip3mu\mu$m are the features with the greatest change with host star spectral type and over geological epochs in the IR \citep[see also][]{rugheimer2013, kaltenegger2007}. The CO$_2$ and O$_3$ features from 9 to 11$\mskip3mu\mu$m are shown in Fig. \ref{ozone} in relative emission and absolute integrated thermal emission flux for 0\% and 60\% cloud cover. In the early-Earth abiotic atmosphere we model a CO$_2$/N$_2$/H$_2$O atmosphere with 0.1$\mskip3mu$bar of CO$_2$. This high CO$_2$ atmosphere shows two strong CO$_2$ features at 9.3 and 10.4$\mskip3mu\mu$m. This feature decreases for the middle two epochs with 0.01$\mskip3mu$bar CO$_2$ and is absent at modern CO$_2$ concentrations. At 3.9$\mskip3mu$Ga, there is not enough abiotic oxygen to produce observable ozone for planets orbiting any stellar type. In general, the O$_3$ feature increases in strength as oxygen rises, first for planets around the hotter star types with more UV and then for all stars by the modern epoch. However, a few notable things happen. The ozone feature is deepest for planets orbiting F stars at 1\% and 10\%$\mskip3mu$PAL O$_2$ and is shallower at 100\%$\mskip3mu$PAL O$_2$ because of a reduced temperature difference between the ozone layer and the surface due to increased ozone heating the stratosphere in the modern atmosphere \citep[see also][]{selsis2000}. At 100\%$\mskip3mu$PAL O$_2$, the hottest F star planet models show the feature weakly in emission \citep[see][]{rugheimer2013}. Also, the ozone feature saturates quickly and so there is a minimal difference between the depth of the feature at 1\% and 10\%$\mskip3mu$PAL O$_2$ for our F0V star planet model. For G, K and M star planet models, the O$_3$ feature increases in depth with increasing planetary age and oxygen concentrations. In the modern atmosphere, at 100\%$\mskip3mu$PAL O$_2$, the O$_3$ feature is strongest between stellar types for planets around K stars due to an interplay of high enough UV to create O$_3$ and a larger temperature difference between the stratosphere and the surface. The wings of the 9.6$\mskip3mu\mu$m O$_3$ feature overlaps with the 9.3$\mskip3mu\mu$m CO$_2$ feature.

\begin{figure}[h!]
\centering
\includegraphics[scale=0.45,angle=0]{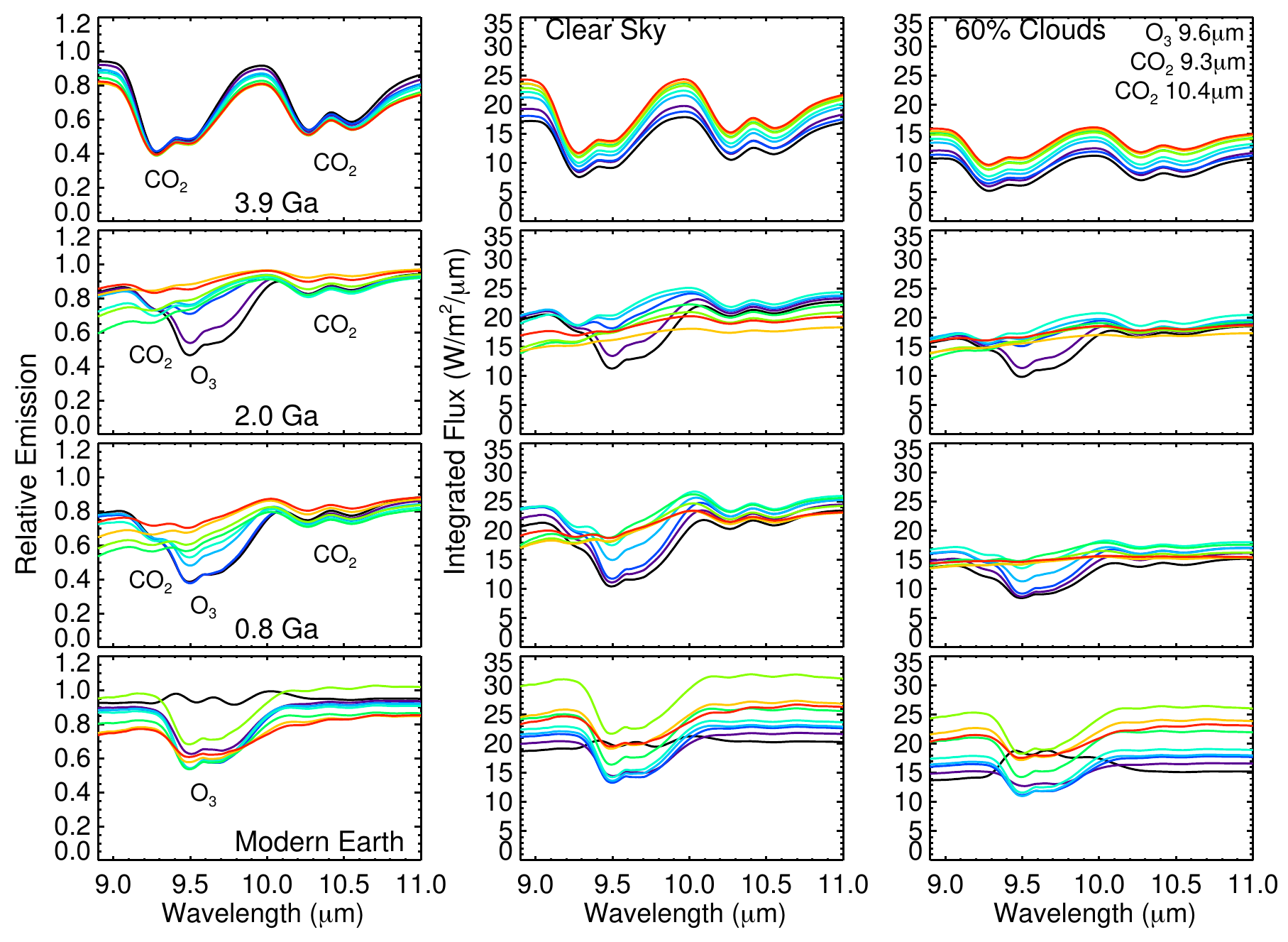}
\caption{Disk-integrated (R =150) spectra of the O$_3$ feature at 9.6$\mskip3mu\mu$m and the CO$_2$ features at 9.3 and 10.4$\mskip3mu\mu$m for clear sky in relative emission (left) and in reflected emergent flux for a 0\% (middle) and a 60\% cloud cover (right). Coloring is same as in Fig. \ref{IRspectra}.\label{ozone}}
\end{figure}

The individual H$_2$O feature at 5-8$\mskip3mu\mu$m  and CH$_4$ feature at 7.7$\mskip3mu\mu$m are shown in relative emission and reflected emergent flux in Fig. \ref{irH2O}. The CH$_4$ feature at 7.7$\mskip3mu\mu$m increases in depth for epochs with higher methane abundance (2.0 and 0.8$\mskip3mu$Ga) and the cooler stars. The water feature from 5-8$\mskip3mu\mu$m is similar in depth for all epochs and is slightly deeper for hot stars due to a larger temperature difference between the absorbing layer and the continuum.

The individual CO$_2$ feature at 15$\mskip3mu\mu$m is shown in relative emission and reflected emergent flux in Fig. \ref{irCO2}. The central plateau observed in the modern Earth-like atmospheres for G and F type stars is due to the hot stratospheres of those stars caused by ozone and as such has been taken as an ``indirect'' ozone feature \citep[see e.g.][]{selsis2000}. We observe no central plateau until ozone concentrations rise to modern Earth levels \citep[see also][]{kaltenegger2007}. The CO$_2$ mixing ratio is constant across all stellar types within each epoch, but the CO$_2$ feature is deepest for the hotter stars in our grid due to the difference in the continuum temperature to the CO$_2$ absorption layer temperature (see Fig. \ref{tpchemsepochs} for temperature vs altitude profiles of the atmospheres). 
\begin{figure}[h!]
\centering
\includegraphics[scale=0.55,angle=0]{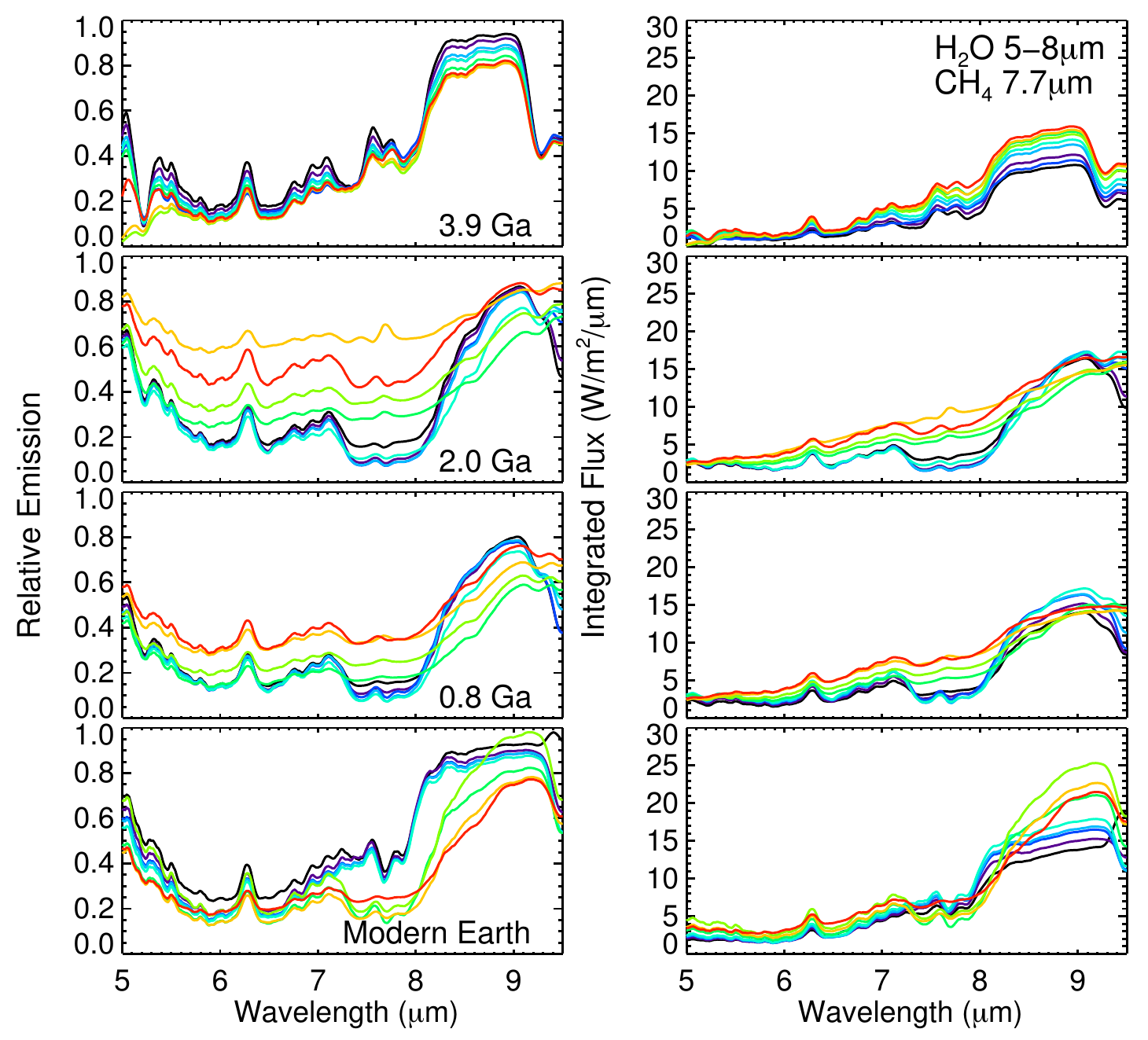}
\caption{Disk-integrated (R =150) spectra of H$_2$O at 5-8$\mskip3mu\mu$m and the CH$_4$ feature at 7.7$\mskip3mu\mu$m for clear sky in relative emission (left) and in reflected emergent flux for a 60\% cloud cover (right). Coloring is same as in Fig. \ref{IRspectra}. \label{irH2O}}
\end{figure}


\begin{figure}[h!]
\centering
\includegraphics[scale=0.55,angle=0]{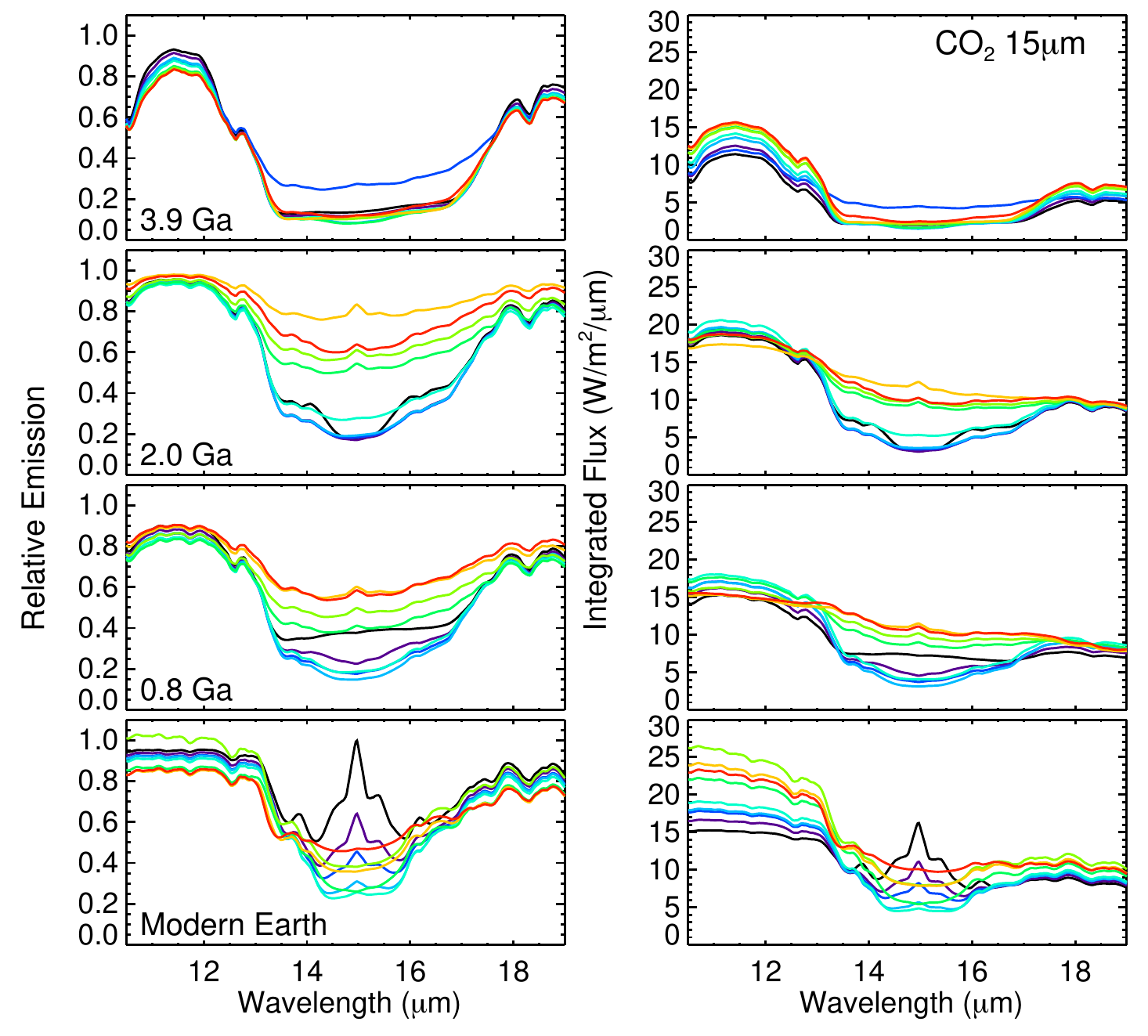}
\caption{Disk-integrated (R =150) spectra of the CO$_2$ feature at 15$\mskip3mu\mu$m for clear sky in relative emission (left) and in reflected emergent flux for a 60\% cloud cover (right). Coloring is same as in Fig. \ref{IRspectra}. \label{irCO2}}
\end{figure}

The strength of spectral features in the IR depends on both the abundance as well as the continuum radiation. For all epochs and all stellar types, H$_2$O, CO$_2$ and CH$_4$ are the dominant features in the IR spectrum. In high CO$_2$ concentration atmospheres, CO is observable in the IR at 4.7$\mskip3mu\mu$m. As oxygen rises, O$_3$ increasingly contributes to the IR spectrum at 9.6$\mskip3mu\mu$m. Note that O$_3$ is a saturated feature, thus it is a very good indicator for even low amounts of oxygen ($\gtrsim$$\mskip3mu$0.01$\mskip3mu$PAL) but does not trace its abundance. N$_2$O and CH$_3$Cl also have features in the IR for epochs where a biological source is assumed. CH$_3$Cl has features at 6-8, 9.8, and 13.7$\mskip3mu\mu$m, and N$_2$O has features at 4.5, 7.8, 8.6, and 17$\mskip3mu\mu$m. Many of these features are obscured by CH$_4$, H$_2$O and CO$_2$. For example, the 17$\mskip3mu\mu$m N$_2$O and the 13.7$\mskip3mu\mu$m CH$_3$Cl features broaden the wings of the 15$\mskip3mu\mu$m CO$_2$ band and would need to be distinguishable for detection. H$_2$O$_2$ at 7.9$\mskip3mu\mu$m and H$_2$CO at 5.7$\mskip3mu\mu$m have their deepest features for the middle two epochs where there is enough oxygen, hydrogen, and carbon species (from the increased O$_2$ and CH$_4$) for their formation but balanced by not too much OH for their destruction. However, the H$_2$O$_2$ and H$_2$CO features are obscured in the full IR spectrum by CH$_4$ and H$_2$O. See Figs. \ref{IRfeaturesSun} to \ref{IRfeaturesM8AHiRes} in the appendix for a plot of the individual features for H$_2$O, CO$_2$, CH$_4$, O$_3$, CO, H$_2$CO, H$_2$O$_2$, N$_2$O, and CH$_3$Cl in the IR for all epochs for the Sun and M8A case.

\section{DISCUSSION}

In the coming decades of exoplanet characterization, we expect to find terrestrial exoplanets at a wide range of stages in their unique geological evolution. Planets will likely have diverse evolutionary paths with perhaps some planets resembling Earth at various points in our own geological history. For an Earth-like atmospheric trajectory, we have simulated planets from prebiotic to current atmosphere based on geological data and presented the emergent exoplanet spectra of these planets.

Detecting the combination of an oxidizing gas and a reducing gas for emergent spectra and secondary eclipse measurements remains the strongest spectroscopic biosignature \citep{lederberg1965, lovelock1975, sagan1993}. Detecting the combination of biosignatures in context of a position in the HZ will aid in reducing the probability of false positive biosignatures. Recent work demonstrates several distinct mechanisms to produce detectable levels of oxygen and ozone through photolysis in CO$_2$-rich planets orbiting M dwarfs \citep{hu2012, tian2014}, around FGKM stars with low-H$_2$/high-CO$_2$ atmospheres \citep{domagal2014}, around pre-main sequence M stars \citep{luger2014}, and in low pressure atmospheres \citep{wordsworth2014}. These studies reinforce the result that individual spectral features should not be used as a biosignature alone and that a strong understanding of the planetary and stellar context will be required for interpretation. And since many of these false positive mechanisms depend on photolysis of CO$_2$ and H$_2$O, measuring the stellar UV radiation environment will be a vital part of future exoplanet characterization efforts.

Several other groups have considered the impact of clouds on spectral features \citep[e.g.][]{desmarais2002, seager2005, tinetti2006a, tinetti2006b, kaltenegger2007, robinson2011, kitzmann2011a, kitzmann2011b, vasquez2013b}. Similar to \citet{kaltenegger2007} and \citet{desmarais2002} we consider the impact of clouds on the spectra by inserting continuum absorbing/emitting layers at the altitudes of the clouds with the cloud fractions represented by a weighted sum of the spectra at each cloud layer. \citet{tinetti2006a, tinetti2006b, robinson2011, vasquez2013b} use explicit scattering and absorbing processes from Mie scattering of liquid water droplets and using parameterized geometrical optical properties of ice clouds. Despite these different cloud implementations, the modeled impact of Earth-like water clouds on spectral features agree well across all groups. As discussed in \citet{kitzmann2011b} and \citet{tinetti2006b}, we find that clouds in the visible can deepen the molecular absorption bands due to the increased reflectance signal from increased albedo and scattered stellar radiation from the cloud deck. However features originating from below the cloud layers in the visible will be suppressed, particularly surface features like the VRE \citep{seager2005, tinetti2006c, rugheimer2013}. In the IR, clouds can dampen features \citep[as seen in][]{kitzmann2011a, vasquez2013b}, though they can also enhance features \citep[e.g.][]{kaltenegger2007, rugheimer2013}. This is due to the IR feature depth depending not only on concentration but on temperature difference between the surface (or cloud layer) and the absorbing or emitting feature.

We considered both clear and cloudy spectral models for the epochs and atmospheres calculated in this paper to show the influence of clouds on the spectra. We have elected to highlight the cloudy atmospheres in the figures as being more relevant to future observations since all known terrestrial planets in our solar system have clouds, though the clear sky models are available publicly in the online database for comparison. The precise cloud type, properties, altitudes, and fractional coverage will also impact the spectral features and is left as future work. 

In the VIS to NIR, the most accessible biosignatures are the O$_2$ feature at 0.76$\mskip3mu\mu$m and one of the multiple CH$_4$ features between 0.6 - 2.4$\mskip3mu\mu$m. The H$_2$O features in the VIS/NIR would provide valuable planetary context of habitability. Detecting the O$_3$ feature at 0.6$\mskip3mu\mu$m will require very high signal to noise observations. The depth and signal of the O$_2$ feature will depend on the global cloud coverage for lower oxygenated atmospheres. Above 10\%$\mskip3mu$PAL, the O$_2$ feature contributes substantial opacity to the directly detectable spectrum (see Fig. \ref{oxygenclouds}). For concentrations of 1\%$\mskip3mu$PAL, detecting the O$_2$ A band will be difficult and will depend on the instrument, resolution, noise, and systematics. Clouds, particularly higher altitude clouds, reduce the depth of the feature, but clouds also boost the reflectivity signal. We note also that 1\%$\mskip3mu$PAL  in the Proterozoic is an optimistic limit, with chromium isotope data suggesting an even lower level of 0.1\%$\mskip3mu$PAL for the mid-Proterozoic \citep[1.8$\mskip3mu$Ga to 0.8$\mskip3mu$Ga, see][]{planavsky2014}. For late M dwarfs, M7V-M9V, the feature is less detectable due to the low level of stellar reflected flux at that wavelength \citep[see also][]{rugheimer2015b}. The CH$_4$ feature is deepest for cooler stars and epochs 2.0 and 0.8$\mskip3mu$Ga.

In the IR, the most accessible biosignatures are the O$_3$ feature at 9.6$\mskip3mu\mu$m and the CH$_4$ feature at 7.7$\mskip3mu\mu$m. The H$_2$O feature at 5-8$\mskip3mu\mu$m and the CO$_2$ features at 9.3, 10.4 and 15$\mskip3mu\mu$m, while not biosignatures themselves directly, would help provide planetary context of habitable conditions since H$_2$O is required for all life as we know it and CO$_2$ and H$_2$O are both greenhouse gases. The central plateau of the CO$_2$ feature indicates a temperature inversion and the presence of an absorbing molecule in the stratosphere such as O$_3$. The 9.6$\mskip3mu\mu$m O$_3$ band is saturated and thus has a non-linear dependence on O$_2$ concentration, making it an excellent qualitative tracer for O$_2$ but not a very good quantitative tracer. The O$_3$ feature at 9.6$\mskip3mu\mu$m will also be in the wings of the CO$_2$ feature at 9.3$\mskip3mu\mu$m depending on the abundance of CO$_2$ in the atmosphere. We find that for F0V, G2V, K7V, and M8V host stars the ozone column depth (cm$^{-2}$) is 2.2, 8.8, 6.5, 9.4 times smaller than for the atmosphere with 1\%$\mskip3mu$PAL O$_2$ compared to the modern atmosphere, respectively (see Table 1 in \citet{rugheimer2015a}). The O$_3$ feature in the IR contributes substantially to the spectrum first for planets orbiting F type stars at 2.0$\mskip3mu$Ga and 0.8$\mskip3mu$Ga. Additionally, for planets around F stars the feature is deepest for these earlier epochs since those stratospheres are not yet hot enough to reduce the temperature contrast between the continuum and the hot stratospheres produced by ozone heating. At 0.8$\mskip3mu$Ga, 10\%$\mskip3mu$PAL O$_2$, the O$_3$ feature contributes to the IR spectrum for planets orbiting G and K stars as well. In the modern atmosphere, the O$_3$ feature is visible for planets orbiting all stellar types, but is strongest for K stars stars due to an interplay of high enough UV to create O$_3$ and a larger temperature difference between the stratosphere and the surface.

For all epochs and stellar types modeled, concentrations of N$_2$O and CH$_3$Cl do not contribute substantially to the emitted spectrum since they overlap with other features and would need high resolution and signal to noise to distinguish them from other features. As such, they would likely be undetectable by the first low resolution and photon limited exoplanet atmosphere characterization missions} \citep[see e.g.][]{selsis2000, kaltenegger2007}. 

\section{CONCLUSIONS}

Our results provide a grid of atmospheric compositions as well as model spectra from the VIS to the IR for possible atmospheres from prebiotic conditions to the modern Earth atmosphere for terrestrial planets at different geological epochs for a grid of F0V-M8V stars ($\teff$ = 7000$\mskip3mu$K to 2400$\mskip3mu$K). These spectra can be a useful input to design instruments and to optimize the observation strategy for direct imaging or secondary eclipse observations with E-ELT or JWST as well as other future mission design concepts such as LUVOIR/HDST. 

Earth-like planets have monotonically increasing planetary albedos with increasing stellar temperature for all grid stars through geological time epochs due to increased Rayleigh scattering. Exoplanet surface temperature varies more complexly with stellar temperature due to efficiency in stratospheric cooling and the column densities of chemical species and available stellar radiation.

At lower biogenic oxygen concentrations of 1\%$\mskip3mu$PAL, corresponding to an Earth-like planet in its paleo to mid-proterozoic epoch, we find that O$_2$ contributes weakly to the spectrum and will require a high resolution and signal to detect at such concentrations. We note that 1\%$\mskip3mu$PAL is an optimistic upper limit of O$_2$ during this epoch and possibly the oxygen level during this time was less than 0.1\%$\mskip3mu$PAL \citep{planavsky2014}. Higher altitude clouds will reduce the depth of the feature but all clouds will increase the reflectivity and signal, see Fig. \ref{oxygen}. Above 10\%$\mskip3mu$PAL O$_2$ corresponding to the neoproterozoic epoch on Earth, the oxygen A band contributes strongly to the detectable emergent spectrum. However, even in modern Earth-like concentrations, O$_2$ is more difficult to detect in direct detection spectra for late M dwarfs due to decreased stellar flux at shorter wavelengths \citep[see][for more details for M dwarf planet models]{rugheimer2015a}. The O$_3$ feature in the IR is less impacted by clouds and the strength is non-linear with oxygen concentration, making that feature much stronger throughout geologic history and for low O$_2$ concentrations. For CO$_2$ concentrations up to 0.01$\mskip3mu$bar, the 9.6$\mskip3mu\mu$m O$_3$ feature may overlap with the wings of the 9.3$\mskip3mu\mu$m CO$_2$ feature. The CO$_2$ plateau at 15$\mskip3mu\mu$m provides an additional diagnostic to determine the presence of an inversion layer. The earliest detection of a combination of biosignatures, i.e. O$_2$/O$_3$ and a reducing species like CH$_4$, is for F stars with 2.0$\mskip3mu$Ga type atmospheres due to both a strong O$_3$ and CH$_4$ feature in the IR.

The model spectra are available by request or at www.cfa.harvard.edu/$\sim$srugheimer/EPOCHspectra/

\section*{ACKNOWLEDGEMENTS}

The authors would like to thank the anonymous reviewer for his/her constructive comments to improve the manuscript. We also thank Claire Cousins and Clara Bl\"attler for useful discussions. This work was supported by a grant from the Simons Foundation (SCOL awards 339489 to SR and 290357 to LK). This work has made use of the MUSCLES M dwarf UV radiation field database. 

\section*{APPENDIX}


\begin{figure*}[ht!]
\centering
\includegraphics[scale=0.65,angle=0]{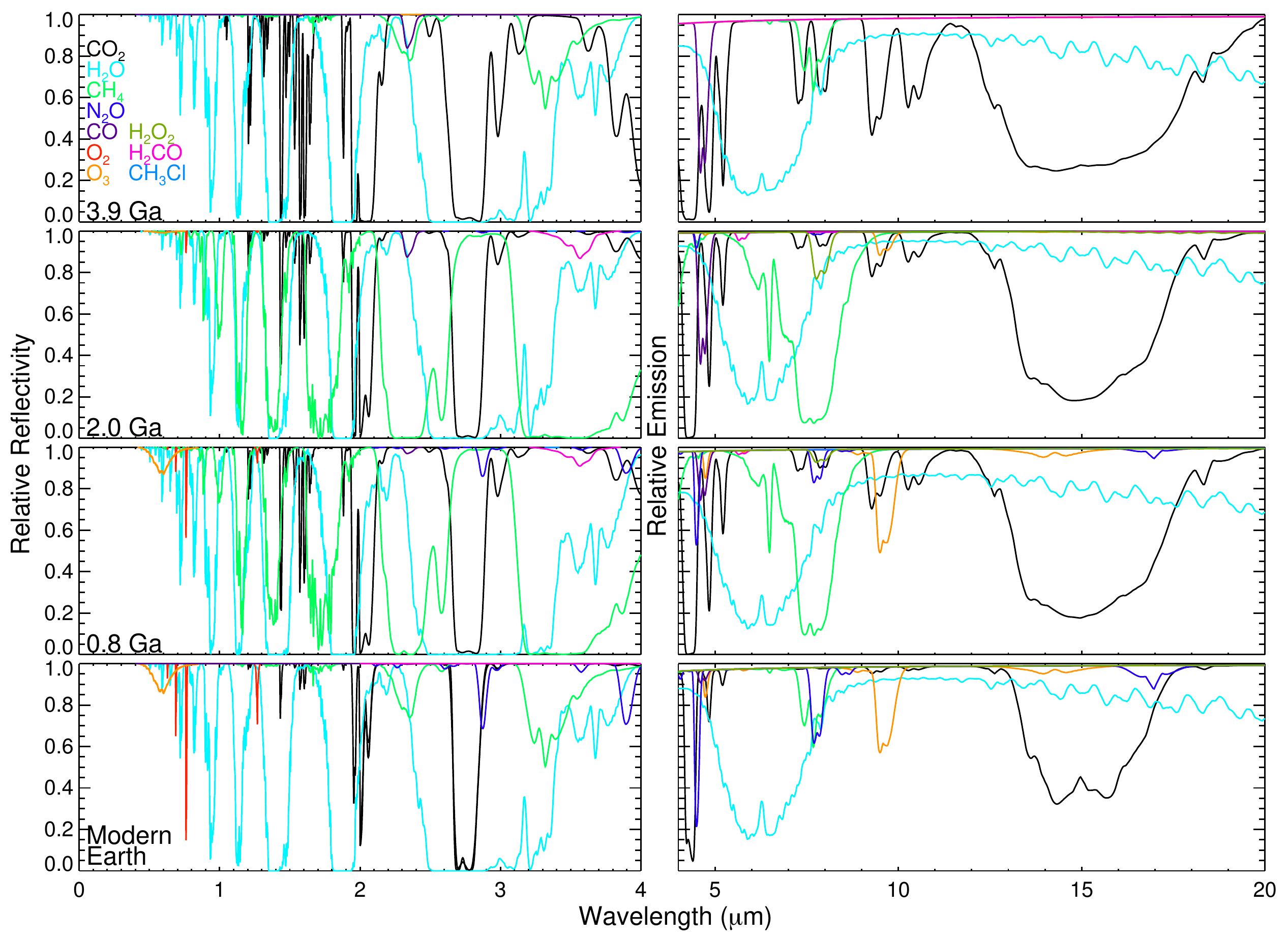}
\caption{Individual spectral components for CO$_2$, H$_2$O, CH$_4$, N$_2$O, CO, O$_2$, O$_3$, H$_2$O$_2$, HNO$_3$, H$_2$CO, and CH$_3$Cl at R = 800 for an Earth-like planet orbiting a solar-analogue stellar model (G2V) for four geological epochs.\label{IRfeaturesSun}}
\end{figure*}

\begin{figure*}[ht!]
\centering
\includegraphics[scale=0.65,angle=0]{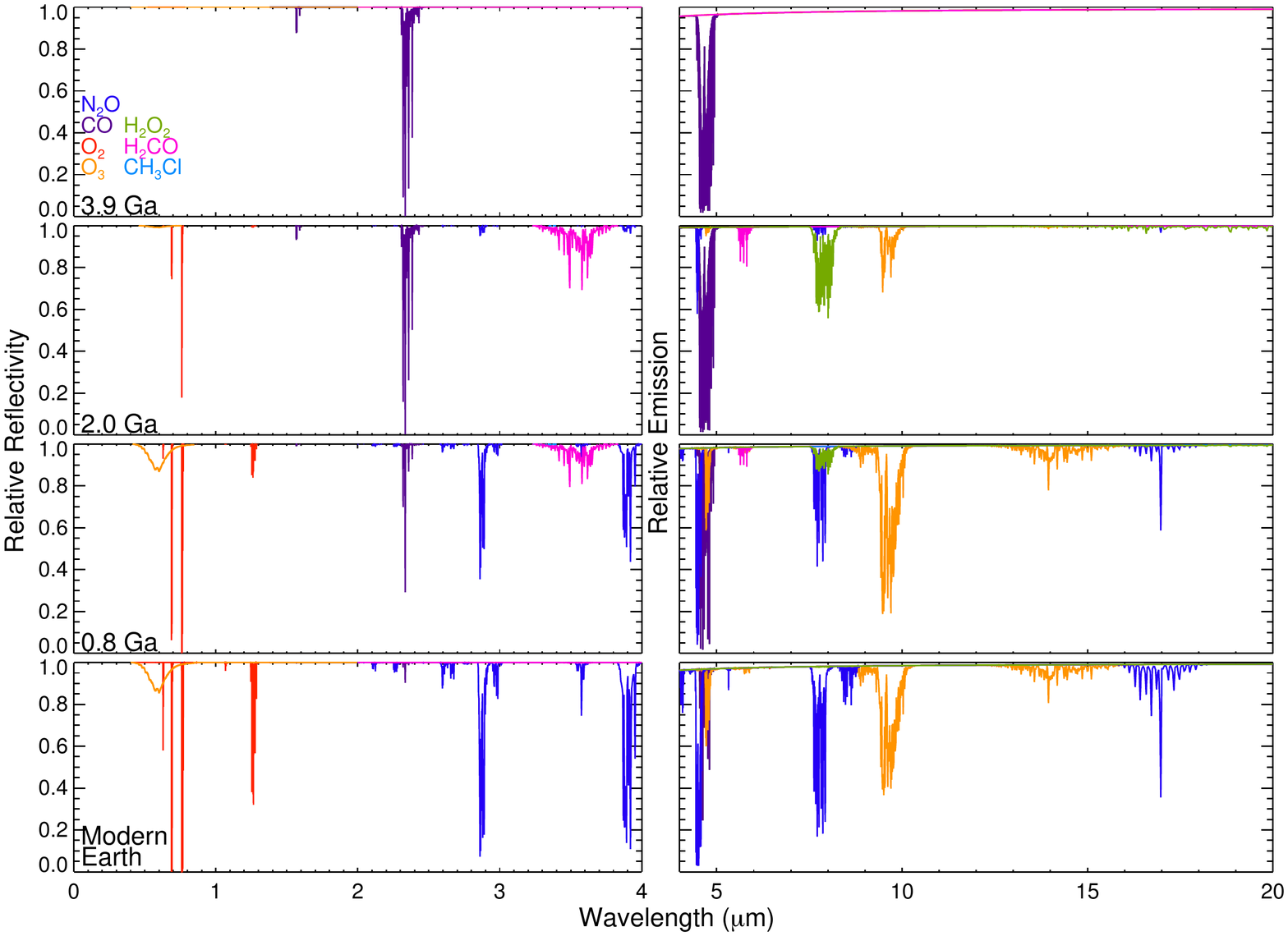}
\caption{Individual spectral components for a subset of chemicals in Fig. \ref{IRfeaturesSun}, for clarity. N$_2$O, CO, O$_2$, O$_3$, H$_2$O$_2$, HNO$_3$, H$_2$CO, and CH$_3$Cl are shown in high resolution (0.1cm$^{-1}$) for an Earth-like planet orbiting a solar-analogue stellar model (G2V) for four geological epochs.\label{IRfeaturesSunHiRes}}
\end{figure*}

\begin{figure*}[ht!]
\centering
\includegraphics[scale=0.65,angle=0]{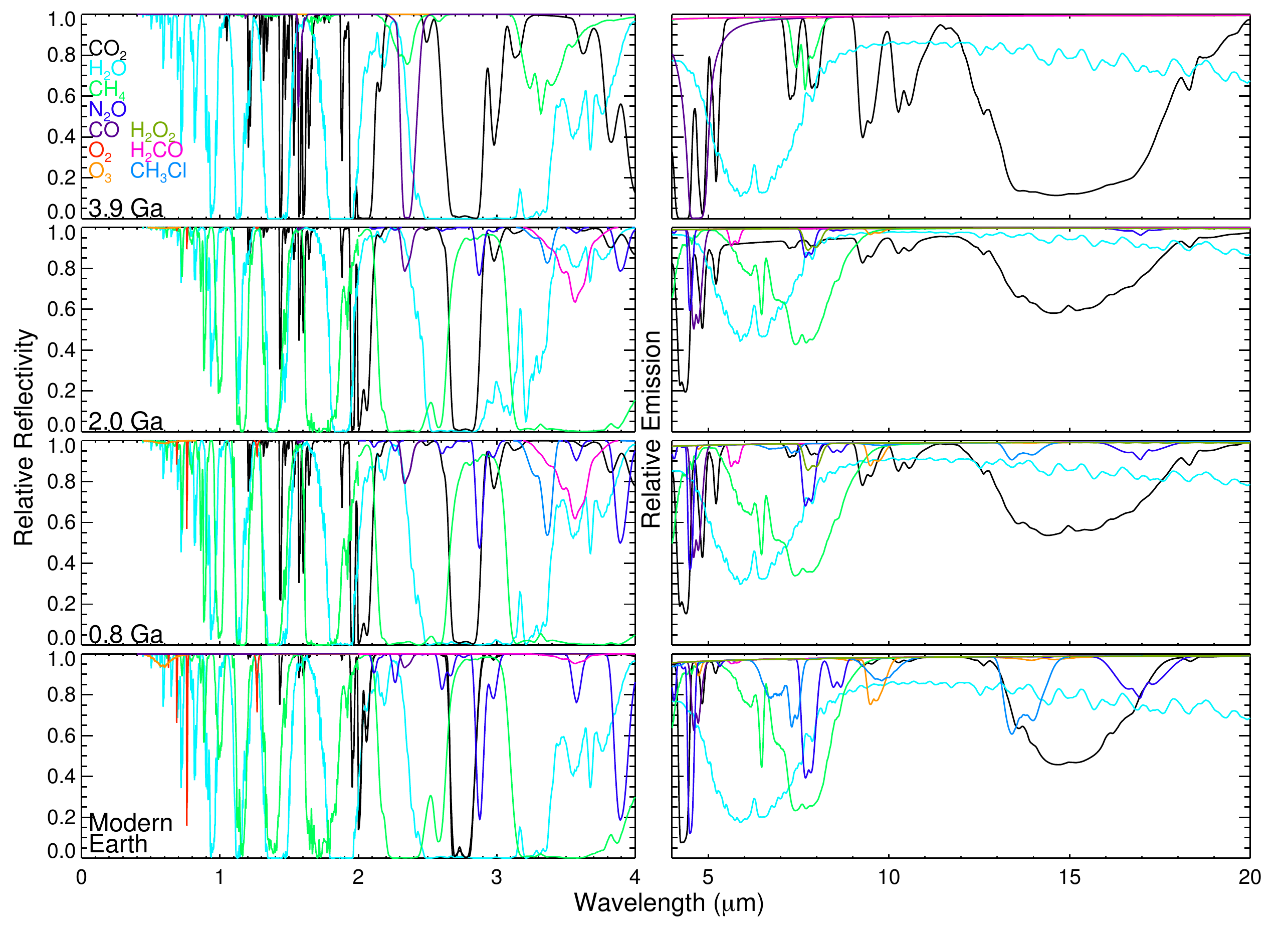}
\caption{Individual spectral components for CO$_2$, H$_2$O, CH$_4$, N$_2$O, CO, O$_2$, O$_3$, H$_2$O$_2$, HNO$_3$, H$_2$CO, and CH$_3$Cl for an Earth-like planet orbiting an M8 active stellar model for four geological epochs.\label{IRfeaturesM8A}}
\end{figure*}

\begin{figure*}[ht!]
\centering
\includegraphics[scale=0.65,angle=0]{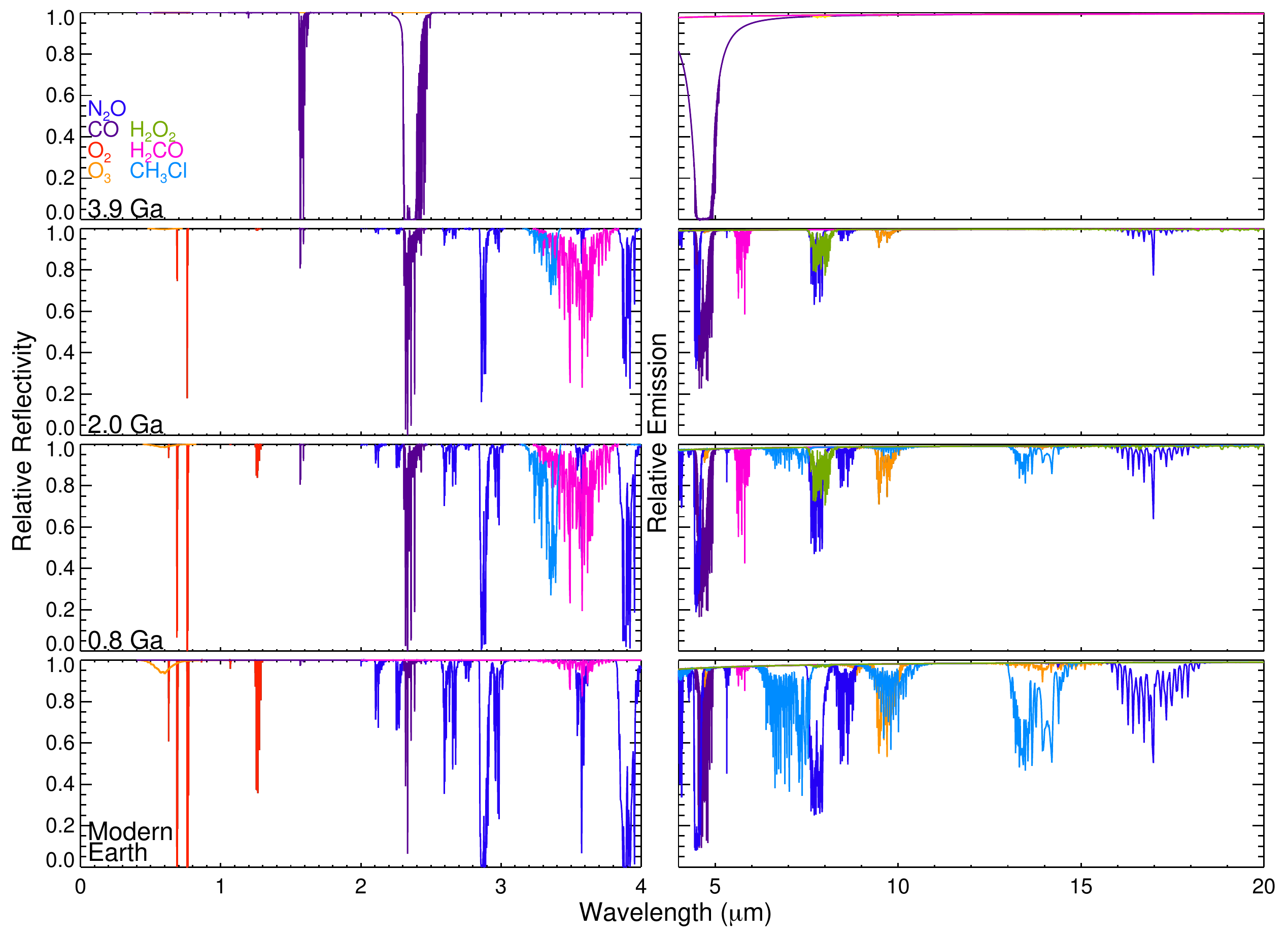}
\caption{Individual spectral components for a subset of chemicals in Fig. \ref{IRfeaturesM8A}, for clarity. N$_2$O, CO, O$_2$, O$_3$, H$_2$O$_2$, HNO$_3$, H$_2$CO, and CH$_3$Cl are shown in high resolution (0.1cm$^{-1}$) for an Earth-like planet orbiting an M8 active stellar model for four geological epochs.\label{IRfeaturesM8AHiRes}}
\end{figure*}

\clearpage


\end{document}